\documentclass[11pt]{llncs}
\usepackage{fullpage}
\usepackage{graphicx}
\usepackage{calc}
\usepackage{xcolor}
\usepackage{colortbl}
\usepackage{amsmath,amssymb} 
\usepackage{minibox}
\usepackage{listings}
\usepackage{times}
\usepackage{picins}
\usepackage{epsfig} 
\usepackage{mathpartir}
\usepackage{bnf}
\usepackage{multirow}
\usepackage{subfigure}
\usepackage{setspace} 
\usepackage{wrapfig}
\usepackage{soul} 
\usepackage[]{todo}
\usepackage{slashbox}
\usepackage{colortbl}
\usepackage{ifthen}

\newcommand{\ignore}[1]{}
\newcommand{\ie}{i.e.,}
\newcommand{\secref}[1]{Sec.~\ref{#1}}
\newcommand{\figref}[1]{Fig.~\ref{#1}}
\newcommand{\defas}{\stackrel{\mathrm{\scriptscriptstyle def}}{=}}
\newcommand{\setof}[1]{\ensuremath{\left \{ #1 \right \}}}
\newcommand{\tuple}[1]{\ensuremath{\left \langle #1 \right \rangle }}
\newcommand{\cO}{{\cal O}}
\newcommand{\cE}{{\cal E}}
\DeclareMathOperator{\traces}{traces}

\newcommand{\sem}[3]{\ensuremath{{[\![#1]\!]}_{#2}^{#3}}}

\DeclareMathOperator{\verdict}{verdict}
\DeclareMathOperator{\controlled}{controlled}
\newcommand{\Data}{\mathrm{Data}}
\newcommand{\func}{\mathit{func}}
\newcommand{\port}{\mathit{port}}
\newcommand{\export}{\mathit{export}}

\newcommand{\guard}{\mathit{guard}}
\newcommand{\source}{\mathit{src}}
\newcommand{\dest}{\mathit{dest}}
\DeclareMathOperator{\var}{var}
\DeclareMathOperator{\insttrans}{inst\_trans}
\DeclareMathOperator{\occur}{occur}
\DeclareMathOperator{\indexf}{index}
\DeclareMathOperator{\used}{used}
\DeclareMathOperator{\monvars}{mon\_vars}
\DeclareMathOperator{\rectrans}{rec\_trans}
\DeclareMathOperator{\selecttrans}{select\_trans}
\DeclareMathOperator{\recvars}{rec\_vars}
\DeclareMathOperator{\rectransi}{rec\_trans-i}
\DeclareMathOperator{\reccomp}{rec\_comp}
\DeclareMathOperator{\recinter}{rec\_inter}
\DeclareMathOperator{\instf}{inst}
\DeclareMathOperator{\inj}{inj}

\newcommand{\enab}{\ensuremath{\mathit{enab}}}

\newcommand{\goesto}[1][]{\stackrel{#1}{\longrightarrow}} 

\newcommand{\toolname}{RE-BIP}

\newcommand{\bfour}{\ensuremath{\mathbb{B}_4}}

\newcommand{\tinit}[1]{{\theta_{{\scriptscriptstyle \mathrm{init}}}^{#1}}}
\newcommand{\trans}[2]{\stackrel{#1}{\longrightarrow}_{#2}}

\newcommand{\Prop}{\mathit{Prop}}
\newcommand{\AP}{\mathit{AP}}
\newcommand{\Safety}{\ensuremath{\mathit{Safety}}}

\newcommand{\true}{\ensuremath{\mathtt{true}}}
\newcommand{\false}{\ensuremath{\mathtt{false}}}

\newcommand{\varsof}[1]{\ensuremath{#1.\mathit{vars}}}
\newcommand{\portsof}[1]{\ensuremath{#1.\mathit{ports}}}
\newcommand{\locsof}[1]{\ensuremath{#1.\mathit{locs}}}
\newcommand{\transof}[1]{\ensuremath{#1.\mathit{trans}}}

\newcommand{\tmp}{{ {\rm \scriptscriptstyle tmp}}}
\newcommand{\instscript}{{ {\rm \scriptscriptstyle inst}}}
\newcommand{\rec}{{ {\rm \scriptscriptstyle rec}}}
\newcommand{\inter}{{ {\rm \scriptscriptstyle inter}}}

\newcommand{\dotport}{\ensuremath{\mathit{port}}}
\newcommand{\dotloc}{\ensuremath{\mathit{loc}}}

\newcommand{\AFF}{{\cal AF}}

\usepackage[belowskip=-5pt,aboveskip=0pt]{caption}
\newboolean{pdf}
\setboolean{pdf}{false}
\begin{document}
\sloppy
\graphicspath{{fig/}}
\title{Runtime Enforcement for Component-Based Systems}
\author{Hadil Charafeddine\inst{1}\and Khalil El-Harake\inst{1}\and Yli\`es Falcone\inst{2}\and Mohamad Jaber\inst{1}}
\institute{
American University of Beirut, Beirut, Lebanon\\
\email{\{hnc01,kme07,mj54\}@aub.edu.lb}
\and
Laboratoire d'Informatique de Grenoble, Universit\'e Grenoble-Alpes, Grenoble, France
\email{Ylies.Falcone@ujf-grenoble.fr}
}
\maketitle
\begin{abstract}
Runtime enforcement is an increasingly popular and effective dynamic validation technique aiming to ensure the correct runtime behavior (w.r.t. a formal specification) of systems using a so-called enforcement monitor.
In this paper we introduce runtime enforcement of specifications on component-based systems (CBS) modeled in the BIP (Behavior, Interaction and Priority) framework. BIP is a powerful and expressive component-based framework for formal construction of heterogeneous systems.
However, because of BIP expressiveness, it remains difficult to enforce at design-time complex behavioral properties.

First we propose a theoretical runtime enforcement framework for CBS where we delineate a hierarchy of sets of enforceable properties (i.e., properties that can be enforced) according to the number of observational steps a system is allowed to deviate from the property (i.e., the notion of $k$-step enforceability).
To ensure the observational equivalence between the correct executions of the initial system and the monitored system, we show that i) only stutter-invariant properties should be enforced on CBS with our monitors, ii) safety properties are $1$-step enforceable.
Given an abstract enforcement monitor (as a finite-state machine) for some $1$-step enforceable specification, we formally instrument (at relevant locations) a given BIP system to integrate the monitor. 
At runtime, the monitor observes and automatically avoids any error in the behavior of the system w.r.t. the specification.
Our approach is fully implemented in an available tool that we used to i) avoid deadlock occurrences on a dining philosophers benchmark, and ii) ensure the correct placement of robots on a map.
\end{abstract}

\section{Introduction}
%
Users wanting to build complex, distributed, heterogeneous systems dispose of a variety of complementary verification techniques such as model-checking, static analysis, testing, and runtime verification to detect bugs and errors.
Techniques are often categorized as static (e.g., model-checking, static analysis) or dynamic (e.g., testing, runtime verification) according to the sort of system information that is analyzed.
Interestingly, these techniques are complementary to each other in terms of desirable features.
For instance, dynamic techniques are scalable (they face the state-explosion problem) and can be applied when some parts of the system are unknown or when verification with other techniques is undecidable.
Both types of techniques take as input some representation of the system, perform some analysis, and yield a verdict indicating the (partial) correctness of the system in addition to providing some form of feedback to the user.
Upon the detection of an error in the system, the user's activity enters a new phase consisting in correcting the system and then submitting the corrected system to the analysis technique.
This process is usually time-consuming and not guaranteed to converge within the time frame associated to system implementation.

\paragraph{Motivations.} We aim at marrying software synthesis and dynamic analysis to solve the aforementioned issue.
While runtime verification complements model-checking, we propose \emph{runtime enforcement} (RE) (cf. \cite{enforceablesecpol,Falcone10,FalconeMFR11}) to complement model repair.
While model repair targets correctness-by-construction, runtime enforcement, as proposed in this paper, targets \emph{correctness-at-operation}.
Runtime enforcement is an increasingly popular and effective dynamic technique aiming at ensuring the correct runtime behavior (w.r.t. a formal specification) of systems using a so-called \emph{enforcement monitor}.
At runtime, the monitor consumes information from the execution (e.g., events) and modifies it whenever it is necessary to comply with the specification by, e.g., suppressing forbidden events.
To the best of our knowledge, enforcing properties at runtime has been only studied for monolithic systems. Moreover, these frameworks remain at an abstract level, and do not specify how systems should be instrumented.

We target component-based systems (CBS) expressed in the BIP (Behavior, Interaction and Priority) framework (see \secref{sec:bip})~\cite{Bliudze-Sifakis-08a,BliudzeS08,bip11}.
BIP uses a dedicated language and toolset supporting a rigorous design flow.
The BIP language allows to build complex systems by coordinating the behavior of a set of atomic components.
Behavior is described with Labelled Transition Systems extended with data and functions written in C.
Coordination between components is layered.
The first layer describes the interactions between components.
The second layer describes dynamic priorities between the interactions to express scheduling policies.
The combination of interactions and priorities characterizes the overall architecture of a system.
This layered architecture confers a strong expressiveness to BIP~\cite{Bliudze-Sifakis-08a}.
Moreover, BIP has a rigorous operational semantics: the behavior of a composite component is formally described as the composition of the behaviors of its atomic components.
This allows a direct relation between the underlying semantic model and its (automatically synthesized) implementation.
\paragraph{Contributions.}
This paper proposes an effective runtime enforcement technique to easily integrate correctness properties into a component-based system.
Our approach favors the design and correctness of safety-critical systems by allowing a separation of concerns for system designers.
Indeed, the functional part of the system and its safety requirements can be designed in separation, and then latter integrated together with our approach.
The resulting supervised system prevents any error from happening.
More specifically, the contributions of this paper are as follows:
\begin{itemize}
\item
to introduce runtime enforcement to monitor and avoid any error in the execution of CBS;
\item
to introduce a new paradigm for runtime enforcement: previous runtime enforcement approaches introduced enforcement monitors that can store ``bad events" in their memory without the possibility of cancelling these events (rolling the system back) to explore alternative executions (see \secref{sec:rw} for a more detailed comparison with related work): the runtime enforcement paradigm proposed in this paper \emph{prevents} the occurrence of misbehaviors in the targeted system;\footnote{In previous RE frameworks, instrumentation of the sysytem is taken for granted.}
\item
to propose an instrumentation technique that minimally alters the behavior of component-based systems and allows the observation and modification of their behavior;
\item
to propose a series of formal transformations that takes as input a component-based system and a desired property to produce a supervised system where the property is enforced: the resulting system produces only the correct executions (of the initial system) w.r.t. the considered property, with low overhead;
\item
to implement the instrumentation and the transformations in RE-BIP, an available toolset;
\item
to validate the effectiveness of the whole approach by enforcing properties over non-trivial systems (where a static hand-coding of the properties using connectors and priorities would have not been tractable): deadlock freedom on dining philosophers and the correct placement of robots on a map. 
\end{itemize}
\paragraph{Challenges.}
When synthesizing enforcement monitors for component-based systems, the main difficulties that arise are:
\begin{itemize}
\item
to handle the possible interactions and synchronizations between components: when intervening on the behavior of a component by e.g., suppressing the execution of a transition, we need to make sure that the synchronized components are also prevented from performing a connected transition;
\item
to preserve the observational equivalence between the initial system (restricted to its correct execution sequences) and the monitored system: for this purpose, i) our transformations leverage the use of priority in BIP, and ii) we identify the set of stutter-invariant properties for which enforcement monitors can be synthesized and integrated into a system while preserving observational equivalence;
\item
to propose an efficient instrumentation technique that ensures that the enforcement monitor receives all events of interest of the property while not degrading the performance of the initial system, for this purpose, the transformations are efficiently implemented in RE-BIP.
\end{itemize}
\paragraph{Paper Organization.}
The remainder of this paper is structured as follows.
Section~\ref{sec:prelim} introduces some preliminaries and notations.
In Section~\ref{sec:bip}, we recall the necessary concepts of the BIP framework.
Section~\ref{sec:re} presents, at an abstract level, a runtime enforcement framework for component-based systems.
Section~\ref{sec:re-bip} shows how to instrument a BIP system to incorporate an enforcement monitor.
Section~\ref{sec:implem} describes RE-BIP, a full implementation of our framework and some benchmarks.
Section~\ref{sec:rw} discusses related work and presents the complementary advantages of our runtime enforcement approach over existing validation techniques. Section~\ref{sec:conclusion} draws some conclusions and perspectives.

\section{Preliminaries and Notation}
\label{sec:prelim}
%
We introduce some preliminary concepts and notations.
\paragraph{Functions and partial functions.} For two domains of elements $E$ and $F$, we note $[E\rightarrow F]$ (resp. $[E\rightharpoondown F]$) the set of functions (resp. partial functions) from $E$ to $F$. When elements of $E$ depend on the elements of $F$, we note $\setof{e\in E}_{f\in F'}$, where $F'\subseteq F$, for $\setof{e\in E\mid f\in F'}$ or $\setof{e}_{f\in F'}$ when clear from context. For two functions $v\in [ X \rightarrow Y]$ and $v'\in [X' \rightarrow Y']$, the substitution function noted $v/v'$, where $v/v' \in [X \cup X' \rightarrow Y \cup Y']$, is defined as: $v/v'(x) = v'(x)$ if $x \in X'$ and $v(x)$ otherwise.
A predicate over some domain $E$ is a function in the set $[E\rightarrow\setof{\true,\false}]$ where $\true$ and $\false$ are the usual Boolean constants. Given, some predicate $p$ over some domain $E$ and some element $e\in E$, we abbreviate $p(e) = \true$ (resp. $p(e)=\false$) by $p(e)$ (resp. $\neg p(e)$).
\paragraph{Sequences.}
Given a set of elements $E$, a sequence of length $n$ over $E$ is denoted $e_1\cdot e_2\cdots e_n$ where $\forall i\in [1,n]: e_i\in E$.
When a sequence $s$ is a prefix of a sequence $s'$, we note it $s\preceq s'$.
When elements of a sequence are assignments, the sequence is delimited by square brackets, e.g., $[x_1:=\mathit{expr_1};\ldots;x_n:=\mathit{expr_n}]$.
Concatenation of assignments or sequences of assignments is denoted by $``;"$.
The set of all sequences over $E$ is noted $E^*$.
\paragraph{Transition Systems.}
Labelled Transition System (LTS) are used to define the semantics of (BIP) systems.
An LTS defined over an alphabet $\Sigma$ is a 3-tuple $\tuple{\mathrm{Lab},\mathrm{Loc},\mathrm{Trans}}$ where $\mathrm{Lab}$ is a set of labels, $\mathrm{Loc}$ is a non-empty set of locations and $\mathrm{Trans}\subseteq \mathrm{Loc}\times \mathrm{Lab} \times \mathrm{Loc}$ is the transition relation.
A transition $\tuple{l,e,l'}\in\mathrm{Trans}$ means that the LTS can move from location $l$ to location $l'$ by consuming label $e$.
We abbreviate $\tuple{l,e,l'}\in\mathrm{Trans}$ by $l\stackrel{e}{\rightarrow}_\mathrm{Trans} l'$ or by $l\stackrel{e}{\rightarrow} l'$ when clear from context.
Moreover, $l\stackrel{e}{\rightarrow}$ is a short for $\exists l'\in\mathrm{Loc}: l\stackrel{e}{\rightarrow} l'$.
The \emph{traces} of LTS $L=\tuple{\mathrm{Lab},\mathrm{Loc},\mathrm{Trans}}$, noted $\traces(L)$, are the finite sequences over $\mathrm{Lab}$ that can be obtained starting from the initial state, concatenating the labels following the transition relation.
\section{BIP - Behavior Interaction Priority}
\label{sec:bip}
%
BIP~\cite{bip11} allows to construct systems by superposing three layers of modeling: Behavior, Interaction, and Priority.
The \emph{behavior} layer consists of a set of atomic components represented by transition systems extended with C functions and data and labeled with communication ports.
The \emph{interaction} layer models the collaboration between components.
The \emph{priority} layer specifies scheduling policies on the interaction layer.
%
\subsection{Component-based Construction}
%
%
%
\subsubsection{Atomic Components.}
An atomic component $B$ is endowed with a finite set of local variables $\varsof{B}$ ranging over a domain $\mathit{\Data}$.
Atomic components synchronize and exchange data with each other through \emph{ports}.
\begin{definition}[Port]
A port $\tuple{p,x_p}$ in atomic component $B$, is defined by a port identifier $p$, and a set of attached local variables $x_p$, where $x_p \subseteq \varsof{B}$.
When clear from context, we denote the port $\tuple{p,x_p}$ via its identifier $p$, and its data variables $x_p$ via the dot notation $\varsof{p}$.
\end{definition}
\begin{definition}[Atomic component]
\label{def:atomic}
An atomic component $B$ is defined as a tuple $\tuple{P,L,T,X,\setof{g_\tau}_{\tau \in T}, \setof{f_\tau}_{\tau \in T}}$, 
where:
\begin{itemize}
  \item $\tuple{P,L,T}$ is an LTS over a set of ports $P$: $L\ignore{=\setof{l_1,l_2,\ldots,l_k}}$ is a set of control locations and $T \subseteq L \times P \times L$ is a set of transitions;
\item $X$ is a set of variables;
\item For each transition $\tau \in T$: 
$g_{\tau}$ is a Boolean condition over $X$: the guard of $\tau$, and
$f_\tau\in \{x := f^x(X)\mid x\in X\}^*$: the computation of $\tau$, a sequence of assignments.
\end{itemize}
\end{definition}
For $\tau = \tuple{l,p,l'}\in T$ a transition of the LTS, $l$ (resp. $l'$) is referred to as the source (resp.
destination) location and $p$ is a port through which an interaction with another component can take place. Moreover, a transition $\tau = \tuple{l,p,l'}\in T$ in the internal LTS involves a transition in the atomic component of the form $\tuple{l,p,g_\tau,f_\tau,l'}$ which can be executed only if the guard $g_\tau$ evaluates to $\true$, and $f_\tau$ is a computation step: a set of assignments to local variables in $X$.

In the sequel we use the dot notation.
Given a transition $\tau = \tuple{l,p,g_\tau,f_\tau,l'}$, $\tau.\source$, $\tau.\port$, $\tau.\guard$, $\tau.\func$, and $\tau.\dest$ denote $l$, $p$, $g_\tau$, $f_\tau$, and $l'$, respectively.
Also, the set of variables used in a transition is defined as $\var(f_\tau) = \setof{x \in X \mid x:= f^x(X) \in f_\tau}$.
Given an atomic component $B$, $\portsof{B}$ denotes the set of ports of the atomic component $B$, $\locsof{B}$ denotes its set of locations, etc.
\begin{figure}[h]
\centering
\scalebox{0.7}{
\ifthenelse{\boolean{pdf}}{
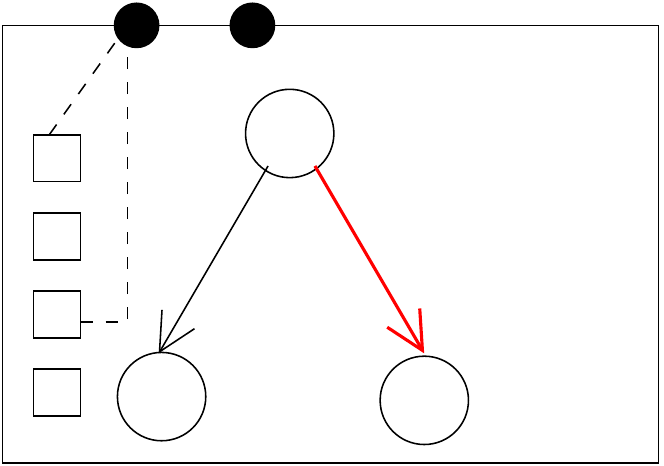
}
{
\input{fig/inst_atomic1.pstex_t}
}
}
\caption{Atomic component\label{fig:atomic}}
\end{figure}
\begin{example}[Atomic component]
Figure~\ref{fig:atomic}, shows the atomic component $\mathit{comp_1}$ with variables $x$, $y$, $z$, and $t$, two ports $p$ and $q$ ($p$ is attached to variables $x$ and $z$), and three control locations $l$, $l'$, and $l''$.
At location $l$, the transition labeled by port $q$ is possible (the guard evaluates to {\true} by default) and the transition labeled by port $p$ is possible provided $x$ is positive. 
When an interaction through $q$ takes place, variable $y$ is assigned to the value of $\mathit{x + t}$.
\end{example}
\begin{definition}[Semantics of atomic components]
  The semantics of atomic component $\tuple{P,L,T,X,\setof{g_{\tau}}_{\tau \in T},\setof{f_{\tau}}_{\tau \in T}}$ is the LTS $\tuple{P,Q,T_0}$, where:
\begin{itemize}
\item $Q = L\times [X\rightarrow \Data]\times (P\cup \{\mathtt{null}\})$,
\item $T_0= \setof{\tuple{\tuple{l,v,p},p'(v_{p'}), \tuple{l',v',p'}}\in Q\times P\times Q\mid \exists \tau = \tuple{l,p',l'} \in T: g_{\tau}(v) \wedge v'=f_{\tau}(v/v_{p'})}$, where $v_{p'} \in[\varsof{p'} \rightarrow \Data]$.
\end{itemize}
\end{definition}
A configuration is a triple $\tuple{l,v,p}\in Q$ where $l \in L$ is a control location, $v \in [X \rightarrow \mathit{\Data}]$ is a valuation of the variables in $X$, and $p \in P$ is the port labeling the last-executed transition or $\mathtt{null}$ when no transition has been executed (i.e., its value is $\mathtt{null}$ at component initialization).
The evolution of configurations $\tuple{l,v,p}\stackrel{p'(v_{p'})}{\rightarrow}\tuple{l',v',p'}$, where $v_{p'}$ is a valuation of the variables in $\varsof{p'}$, is possible if there exists a transition $\tuple{l,p',g_\tau,f_\tau,l'}$, s.t. $g_\tau(v)=\true$.
As a result, the valuation $v$ of variables is modified to $v'=f_\tau(v/v_{p'})$.
%
\subsection{Creating Composite Components}
%
Assuming some atomic components $B_1,$ $\ldots,B_n$, we show how to connect the components in the set $\{B_i\}_{i\in I}$ with $I\subseteq [1,n]$ using a \emph{connector}.

A connector $\gamma$ is used to specify possible interactions, {\ie} the sets of ports that have to be jointly executed.
Two types of ports (\textit{synchron}, \textit{trigger}) are defined in order to specify the feasible interactions of a connector.
A \textit{trigger} port (represented by a triangle) is active: the port can initiate an interaction without synchronizing with other ports.
A \textit{synchron} port (represented by a circle) is passive: the port needs synchronization with other ports to initiate an interaction.
A feasible interaction of a connector is a subset of its ports s.t. either it contains some trigger, or it is maximal.

\parpic(5cm,0.8cm)[r]{
\scalebox{0.9}{
\ifthenelse{\boolean{pdf}}
{\includegraphics{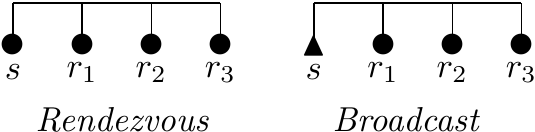}}
{\includegraphics{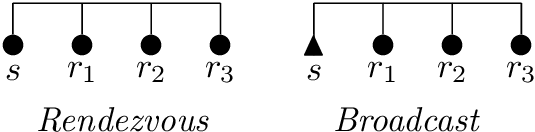}}
}
}
On the right two connectors are depicted: \textit{Rendezvous} (only the maximal interaction $\setof{s,r_1,r_2,r_3}$ is possible), \textit{Broadcast} (all interactions containing trigger port $s$ are possible).
\begin{definition}[Connector]
\label{def:connector}
A connector $\gamma$ is a tuple $\tuple{{\cal P}_\gamma,t,G,F}$, where:
\begin{itemize}
\item ${\cal P}_\gamma = \setof{p_i\mid p_i\in B_i.P}_{i \in I}$ s.t. $\forall i\in I: {\cal P}_\gamma \cap B_i.P = \setof{p_i}$,
\item $t\in [{\cal P}_\gamma\rightarrow \{\true,\false\}]$ s.t. $t(p)=\true$ if $p$ is trigger (and $\false$ otherwise),
\item $G$ is a Boolean expression over the set of variables $\cup_{i\in I}\ \varsof{p_i}$ (the guard),
\item $F$ is an update function defined over the set of variables $\cup_{i\in I}\ \varsof{p_i}$.
\end{itemize}
\end{definition}
${\cal P}_\gamma$ is the set of connected ports of $\gamma$.
A port in ${\cal P}_\gamma$ is tagged using a function $t$ indicating whether it is a trigger or synchron. Moreover, for each $i\in I$, $\varsof{p_i}$ is a set of variables associated with port $p_i$.

A communication between the atomic components of $\{B_i\}_{i\in I}$ through a connector $({\cal P}_\gamma, t, G,F)$ is defined using the notion of \emph{interaction}.
\begin{definition}[Interaction]
\label{def:int}
A set of ports $a=\setof{p_j}_{j\in J} \subseteq {\cal P}_\gamma$ for some $J\subseteq I$ is an interaction of $\gamma$ if either
there exists $j \in J$ s.t. $p_j$ is trigger, or,
for all $j \in J$, $p_j$ is synchron and $\setof{p_j}_{j\in J} = {\cal P}_\gamma$.
\end{definition}
An interaction $a$ has a guard and two functions $G_a,F_a$, respectively obtained by projecting $G$ and
$F$ on the variables of the ports involved in $a$.
We denote by ${\cal I}(\gamma)$ the set of interactions of $\gamma$ and ${\cal I}(\gamma_1)\cup\ldots\cup {\cal I}(\gamma_n)$ by ${\cal I}(\gamma_1,\ldots,\gamma_n)$.
Synchronization through an interaction involves two steps: evaluating $G_a$, and applying the update function $F_a$. 
%
\begin{definition}[Composite component]
A composite component is defined from a set of available atomic components $\setof{B_i}_{i\in I}$ and a set of connectors $\Gamma$.
The connection of the components in $\setof{B_i}_{i\in I}$ using the set $\Gamma$ of connectors is denoted by $\Gamma(\setof{B_{i}}_{i\in I})$.
\end{definition}
Note that a composite component obtained by composing a set of atomic components can be composed with other components in a hierarchical and incremental fashion using the same operational semantics.
\begin{definition}[Semantics of composite components]
\label{def-runtimesemanticscomposite}
A state $q$ of a composite component $\Gamma(\{B_1, \ldots, B_n\})$, where $\Gamma$ connects the $B_i$'s for $i\in [1,n]$, is an $n$-tuple $q=\tuple{q_1,\ldots,q_n}$ where $q_i=\tuple{l_i,v_i,p_i}$ is a state of $B_i$. Thus, the semantics of $\Gamma(\{B_1, \ldots, B_n\})$ is defined as a transition system $\tuple{Q,A,\goesto}$, where:
\begin{itemize}
\item $Q= B_1.Q\times \ldots\times B_n.Q$, 
\item $A = \cup_{\gamma \in \Gamma}\setof{a \in {\cal I}(\gamma)}$ is the set of all possible interactions,
\item $\goesto$ is the least set of transitions satisfying the following rule: 
\begin{mathpar}
\inferrule*
{
  \exists\gamma \in \Gamma: \gamma = \tuple{P_\gamma,t,G,F} \and \exists a \in {\cal I}(\gamma): a = \setof{p_i}_{i \in I} \and
    G_a(v(X)) \hva\\
    \forall i\in I:\ q_i \goesto[p_i(v_i)]_i q'_i \wedge v_i = F_{a_i}(v(X)) \and
    \forall i\not\in I:\ q_i = q'_i
}
{
  \tuple{q_1,\dots,q_n} \goesto[a] \tuple{q'_1,\dots,q'_n}
}
\end{mathpar}
where $X$ is the set of variables attached to the ports of $a$, $v$ is the global valuation, and $F_{a_i}$ is the partial function derived from $F$ restricted to the variables of $p_i$.
\end{itemize}
\end{definition}
The meaning of the above rule is the following: if there exists an interaction $a$ s.t. all its ports are enabled in the current state and its guard ($G_a(v(X))$) evaluates to \true, then the interaction can be fired. When $a$ is fired, all involved components evolve according to the interaction and not involved components remain in the same state. 

Several distinct interactions can be enabled at the same time, thus introducing non-determinism in the product behavior.
Priorities can reduce non-determinism: one of the interactions with the highest priority is chosen in a non-deterministic manner.
\begin{definition}[Priority]
  \label{defn:priority}
  Let $C = \tuple{Q,A,\goesto}$ be the behavior of the composite component $\Gamma(\setof{B_1, \ldots, B_n})$.  
  A {\em priority model} $\pi$ is a
  strict partial order on the set of interactions $A$.
  We  abbreviate $\tuple{a,a'}\in \pi$ by $a \prec_\pi a'$ or $a \prec a'$ when clear from the context. Adding priority model $\pi$ over $\Gamma(\setof{B_1,\ldots,B_n})$ defines a new composite component $\pi\big(\Gamma(\setof{B_1,\ldots,B_n})\big)$ noted $\pi(C)$ and whose behavior is defined by $\tuple{Q,A,\goesto_\pi}$, where $\goesto_\pi$ is the least set of transitions satisfying the following rule: 
\[
\inferrule*
	{
      q \goesto[a] q' \and
      \neg\big(\exists a'\in A,\exists q''\in Q: a \prec a' \wedge q \goesto[a'] q'' \big)
    }
    {
      q \goesto[a]_\pi q'
    }
\]
\end{definition}
An interaction $a$ is enabled in $\pi(C)$ whenever $a$ is enabled in $C$ and $a$ is maximal according to $\pi$ among the enabled interactions in $C$.

We adapt the notion of \emph{maximal progress} to BIP systems. In BIP, the maximal progress property is expressed at the level of connectors. For a given connector $\gamma$, if one interaction $a \in {\cal I}(\gamma)$ is contained in another interaction $a' \in {\cal I}(\gamma)$, then the latter has a higher priority, unless there exists an explicit priority stating the contrary. Maximal progress is enforced by the BIP engine.
\begin{definition}[Maximal progress]
\label{def:maximalprogress}
Given a connector $\gamma$ and a priority model $\pi$, we have: $\forall a,a' \in {\cal I}(\gamma)$: $(a \subset a') \wedge (a' \prec a \notin \pi) \implies a \prec a'$.
\end{definition}
Finally, we consider systems defined as a parallel composition of components together with an initial state.
\begin{definition}[System]
\label{def:system}
A BIP system ${\cal S}$ is a pair $\tuple{B,\mathit{Init}}$ where $B$ is a component and $\mathit{Init}\in B_1.L\times \ldots\times B_n.L$ is the initial state of $B$.
\end{definition}
For the sake of simpler notation, $\mathit{Init}$ designates both the initial state of the system at the syntax level and the initial state of the underlying LTS.
\paragraph{Hierarchical connectors~\cite{Bliudze-Sifakis-08a}.} 
Given a connector $\gamma$ we denote by $\gamma.\export$ to be the exported port of connector $\gamma$, which is used to build hierarchical connectors.
In that case, we use upward and downward update functions instead of update functions only.
\parpic(2.2cm,1.4cm)[r]{
\scalebox{1}{
\ifthenelse{\boolean{pdf}}
{\includegraphics{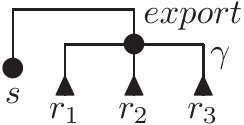}}
{\includegraphics{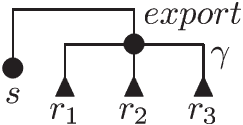}}
}
}
On the right-hand side, we show an example of hierarchical connectors.
All interactions containing $s$ and an interaction of $\gamma$ are possible, i.e., $\setof{s r_1, s r_2, s r_3, s r_1 r_2, s r_1 r_3, s r_2 r_3, s r_1 r_2 r_3}$. 
We will use hierarchical connectors to connect the monitor with the instrumented system in an elegant manner (see \secref{sec:re-bip}).
%

\section{A Runtime Enforcement Framework for Component-Based Systems}
\label{sec:re}
%
We propose an abstract runtime enforcement framework specific to CBS.
Compared to previous runtime enforcement frameworks for monolithic systems, our framework i) takes into account how we instrument CBS to incorporate monitors along with their enforcement abilities, and ii) introduces a hierarchy of enforceable properties.
We shall define how properties are specified, what is the hierarchy of enforceable properties for CBS, what are enforcement monitors, and, what it means for an enforcement monitor to enforce a property.
\paragraph{Preliminaries.}
We consider that the specification of interest is modeled as a property over an alphabet of relevant system events $\Sigma$.
A property $\Pi$ over $\Sigma$ is a subset of $\Sigma^*$.
If a sequence $\sigma$ belongs to a property $\Pi$, we note it $\Pi(\sigma)$.
To evaluate sequences of system events against properties, we shall use the  truth-domain $\bfour$ containing  the truth values true ($\top$), false ($\bot$), currently true ($\top_c$), and currently false ($\bot_c$)~\cite{LeuckerBS08JLC,DBLP:conf/rv/FalconeFM09}.
Given a sequence $\sigma\in\Sigma$ and a property $\Pi\subseteq\Sigma^*$, the evaluation of $\sigma$ against $\Pi$~\cite{DBLP:conf/rv/FalconeFM09} is given by function $\sem{\cdot}{\bfour}{\Pi}$, and defined as:
$
\sem{\sigma}{\bfour}{\Pi}
= \left\{
\begin{array}{ll}
\top & \text{if } \Pi(\sigma) \wedge \forall \sigma'\in\Sigma^*: \Pi(\sigma\cdot\sigma'),\\
\top_c & \text{if } \Pi(\sigma) \wedge \exists \sigma'\in\Sigma^*: \neg \Pi(\sigma\cdot\sigma'),\\
\bot_c & \text{if } \neg\Pi(\sigma) \wedge \exists \sigma'\in\Sigma^*: \Pi(\sigma\cdot\sigma'),\\
\bot & \text{if } \neg\Pi(\sigma) \wedge \forall \sigma'\in\Sigma^*: \neg\Pi(\sigma\cdot\sigma').
\end{array}
\right.
$

We consider \emph{safety} properties which specify that nothing bad should ever happen.\footnote{Here, without restriction, we assume that $\epsilon\in \Safety(\Sigma)$, otherwise enforcement monitors have no chance to enforce the desired property.}
The set of safety properties over $\Sigma$ is noted $\Safety(\Sigma)$.
Safety properties are the prefix-closed properties of $\Sigma^*$: $\Pi\in\Safety(\Sigma)$ iff $\forall \sigma\in\Sigma^*: \Pi(\sigma) \implies \forall \sigma'\preceq\sigma: \Pi(\sigma')$.
Note, for safety properties, only three truth-values are needed for the evaluation of sequences with function $\sem{\cdot}{\bfour}{\Pi}$, i.e., $\forall \Pi\in\Safety(\Sigma), \forall \sigma\in\Sigma^\ast: \sem{\sigma}{\bfour}{\Pi} \in\{\bot, \top_c,\top\}$.
%
\subsection{Specifying Properties of Component-Based Systems~\cite{FalconeJNBB13}}
\label{sec:re:spec}
%
We consider state-based specifications to express desired behaviors.
To be general, we only describe the events of the specification language.
We consider events built as Boolean expressions over a set of atomic propositions.
Atomic propositions express conditions on the local information of components.
For instance, an atomic proposition can express a condition on the lastly executed port, the current locations of a components, the values of variables in different components, etc.
(e.g., ``in component $B_1$, variable $x$ should be positive if in component $B_2$ variable $y$ is negative'').
More formally, an event of $\pi(C)$ is defined as a state formula over the atomic propositions expressed on components involved in $\pi(C)$. Let $\AP$ denote the set of atomic propositions defined with the following grammar (where $*\in\{=,\leq\}$):
\begin{tabular}{rclcrcl}
\multicolumn{1}{r}{Atom} & \multicolumn{1}{c}{$::=$}& \multicolumn{5}{l}{$\text{cpnt}_1.\text{var}_1 * \text{cpnt}_2.\text{var}_2$ $\mid$ $\text{cpnt}.\text{var} * \text{a\_val}$ $\mid$  $\text{cpnt}.\text{loc} = \text{a\_loc} $ $\mid$ $\text{cpnt}.\text{port} = \text{a\_port} $}\\
cpnt.var &$::=$& $x \in \cup_{i\in [1,n]} \varsof{B_i}$ & ~ & $\text{a\_val}$ &$::=$& $v\in \Data$
\end{tabular}

\begin{tabular}{rclcrcl}
  $\text{a\_loc}$ &$::=$& $s\in \cup_{i\in [1,n]} \locsof{B_i}$ & & $\text{a\_port}$ &$::=$& $p\in \cup_{i\in [1,n]} \portsof{B_i}$
\end{tabular}

An atomic proposition compares the values of some variables, the current location, or the port that is on the last executed transition.
Let $\Sigma$ denote the set of events defined as Boolean combinations of atomic propositions.
The property $\Pi$ of interest (over $\Sigma$) will be specified through its runtime oracle, a finite-state machine over $\Sigma$ (see \secref{sec:re-cbs-abstract}).
%
In the sequel, we suppose that all atomic propositions appearing in the property affect its truth-value\footnote{Otherwise, some simplification of the specification shall be performed beforehand. For instance, such simplification should rule out events of the form $a \vee \neg a$ where $a\in \text{Atom}$.}
We use $\Prop: \Sigma \rightarrow 2^{AP}$ for the set of atomic propositions used in an event $e\in \pi(C)$. 
%
For $\mathit{ap}\in \Prop(e)$, $\used(ap)$  is the sequence of pairs formed by the components and the variables (or locations or
ports) that are used to define $\mathit{ap}$. The expression $\used(\mathit{ap})$ is defined using a \emph{pattern-matching} as follows:
\[
\begin{array}{rcl}
 \mathtt{used(ap)}  & = & \mathtt{match (ap)} \;\mathtt{with}\\
        & \mid &  \mathtt{cpnt_1.var_1 * cpnt_2.var_2 \rightarrow (cpnt_1,var_1) \cdot (cpnt_2,var_2)} \\
        & \mid &  \mathtt{cpnt.var * val \rightarrow (cpnt,var)}\\
        & \mid &  \mathtt{cpnt.loc = a\_loc \rightarrow (cpnt,loc)}\\
        & \mid &  \mathtt{cpnt.port = a\_port \rightarrow (cpnt,port)}
\end{array}
\]
%
\subsection{Enforceable Properties on Component-based Systems}
\label{sec:re:enforceable}
%
Two constraints will delineate the set of enforceable properties:\footnote{Contrarily to other runtime enforcement frameworks such as~\cite{enforceablesecpol,RuntimeNonSafety}, we do not consider specifications over infinite sequences but finite sequences.
It avoids dealing with enforceability issues due to the semantics of the specification formalism (over infinite sequences, see~\cite{FalconeFM12} for a detailed explanation).
In that case, for monolithic systems, all properties are enforceable.}
\emph{$k$-step tolerance} and \emph{stutter-invariance}.
These constraints will be justified at a technical level in \secref{sec:re-bip}.
%
\paragraph{$k$-step tolerance and enforceability.}
$k$-step tolerance represents the maximal number of steps for which the system can deviate from the property and can still roll back.
This might be due to the criticality of the system or the controlability endowed to our enforcement monitors on the system.
Moreover, when an enforcement monitor intervenes in the system (to roll it back to a previous state), it should not destroy any (future) correct behavior.
That is, a monitor has to be able to determine that a deviation is definitive at some point.
In other words, on any execution sequence, if the last events made the property unsatisfied, then after some steps, on receiving an event the monitor should be able to determine that there is no possible future behavior s.t. the execution again becomes correct. 
It is thus legitimate for the monitor to intervene.
\begin{definition}[$k$-step enforceability]
\label{def:kstep-enforceable}
$\Pi$ is enforceable with $k$ memorization steps, or $k$-step-enforceable, if
$
\max \big\{ |\sigma| \mid \exists \sigma'\in\Sigma^*: \sem{\sigma'}{\bfour}{\Pi}=\top_c \wedge \forall\sigma_p \preceq \sigma: \sem{\sigma'\cdot \sigma_p}{\bfour}{\Pi}=\bot_c \big\} < k.
$
The set of $k$-step enforceable properties over $\Sigma$ is noted $\mathit{Enf(k,\Sigma)}$.
\end{definition}
A property $\Pi\subseteq\Sigma^\ast$ is $k$-step-enforceable, if the length of its maximal factor $\sigma$ for which there exists a sequence $\sigma'$ (without the factor) that evaluates to $\top_c$ and all sequences $\sigma'\cdot\sigma_p$ obtained by appending a prefix $\sigma_p$ of $\sigma$ to $\sigma'$ evaluate to $\bot_c$.
The constant $k$ additionally represents the maximal ``roll-back distance" of enforcement monitors, i.e., the number of observational steps, an enforcement monitor can revert the system.
\begin{proposition}[A hierarchy of enforceable properties]
There exists a hierarchy of enforceable properties in the sense of Definition~\ref{def:kstep-enforceable} where:
\begin{enumerate}
\item
$\forall k,k'\in\mathbb{N}: k\leq k' \implies \mathit{Enf(k,\Sigma)} \subseteq \mathit{Enf(k',\Sigma)}$;
\item
for regular properties, $k$-step enforceability is decidable.
\end{enumerate}
\end{proposition}
As the first endeavor in introducing runtime enforcement for CBS, we consider enforcement monitors that have the ability to memorize \emph{one} state of the system and thus restore the system up to one observational step in the past.\footnote{The complexity of the instrumentation depends on the number of steps one wants to be able to roll-back the system (see \secref{sec:re-bip}). Considering more than one step is left for future work.}
\begin{proposition}
\label{prop:enforceable-safety}
All safety properties are $1$-step-enforceable as per Definition~\ref{def:kstep-enforceable}: $\Safety(\Sigma) \subseteq \mathit{Enf(1,\Sigma)}$.
\end{proposition}
Safety properties are prefix-closed languages.
Thus when our monitors detect a deviation from the normal behavior on one event, it is legitimate for them to intervene because all deviations from the normal behavior are definitive.

%
\paragraph{Stutter-invariance.}
Stutter-invariance~\cite{Lamport83,WilkeTemporal} is a classical notion of concurrent systems.
Imposing stutter-invariance of specifications stems from how it is required to instrument component-based systems to allow enforcement monitoring.

As seen in \secref{sec:re:spec}, properties are built over atomic propositions which depend on the lastly executed port, the current location of a component, or the values of variables.
Thus our monitor should be able to observe any change in the system that can impact the satisfaction of an atomic proposition.
Since our monitors should be able to revert the global state of a system one step in the past, and as we shall see in \secref{sec:re-bip}, instrumenting a transition in a component implies to instrument all transitions synchronized (through a port/interaction) with that transition.
This is a consequence of BIP semantics (see Definition~\ref{defn:priority}).
Note that, even if an instrumented transition does not interfere with variables observed by the monitor, it is necessary to instrument it for recovering purposes.
Those transitions might be synchronized with other transitions through some interactions.
In that case, when executing one of these (instrumented) interactions, the monitor receives the same ``event" while the system has not changed.
The evaluation of the property w.r.t. the input sequence of events should not change.
Such requirement imposes that the considered properties are stutter-invariant.
\begin{definition}[Stutter-invariance~\cite{Lamport83,WilkeTemporal}]
  Two sequences of events $\sigma,\sigma'\in\Sigma^*$ are stutter-equivalent if there exist $a_0,\ldots,a_k\in\Sigma$ for some $k$ s.t. $\sigma$ and $\sigma'$ belong to the set defined by the regular expression $a_0^+\cdot a_1^+\cdots a_k^+$.
A property $\Pi\subseteq\Sigma^*$ is stutter-invariant, if for any stutter-equivalent sequences $\sigma, \sigma'\in\Sigma^*$, we have ($\sigma\in\Pi$ and $\sigma'\in\Pi$) or ($\sigma\notin\Pi$ and $\sigma'\notin\Pi$).
\end{definition}
Based on Proposition~\ref{prop:enforceable-safety}, we finally consider the set of \emph{stutter-invariant} safety properties as the enforceable properties on component-based systems.
\begin{remark}
Determining whether a property is stutter-invariant is decidable for regular properties using an automata-based representation~\cite{WilkeTemporal}.
Determining whether a property is a safety property is obviously decidable for regular properties.
For these purposes, the automata-based representation of the property is the monitor.
\end{remark}
%
%
\subsection{Runtime Enforcement for Component-based Systems, at an Abstract Level}
\label{sec:re-cbs-abstract}
%
%
We formalize runtime oracles (input to our enforcement framework), enforcement monitors, and how the latter enforce a property described by a runtime oracle.
\paragraph{Runtime oracle.}
A runtime oracle is a finite-state machine that consumes events from the system and produces verdict on each received event.
\begin{definition}[Runtime oracle~\cite{FalconeFM12}]
\label{def:oracle}
An {\em oracle} $\cO$ is a tuple $\tuple{\Theta^\cO,\tinit\cO,\Sigma,\trans{}\cO,\bfour,\verdict^\cO}$.
The finite set $\Theta^\cO$ denotes the control states and $\tinit\cO \in \Theta^\cO$ is the initial state.
The complete function $\trans{}{\cO} : \Theta^\cO \times \Sigma \rightarrow \Theta^\cO$ is the transition function.
In the following we abbreviate $\trans{}{\cO}(\theta, a)=\theta'$ by $\theta \trans{a}{\cO} \theta'$.
Function $\verdict^\cO:\Theta^\cO\rightarrow \bfour$ is an output function, producing verdicts ({\ie} truth-values) in $\bfour$ from control states.
\end{definition}
\begin{wrapfigure}{r}{4cm}
\centering
\ifthenelse{\boolean{pdf}}
{\includegraphics[scale=0.8]{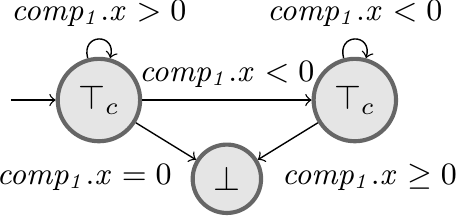}}
{\includegraphics[scale=0.8]{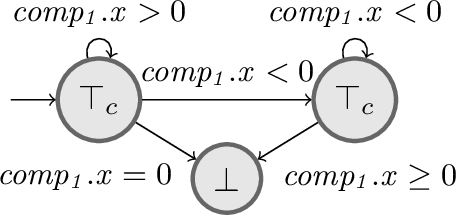}}
\caption{Runtime oracle}
\label{fig:runtime-oracle}
\end{wrapfigure}
Runtime oracles are independent from any formalism used to generate them and are able to check any linear-time property~\cite{DBLP:conf/rv/FalconeFM09}.\footnote{The runtime oracle to be synthesized from a specification, using some monitor synthesis algorithm.
We assume the oracle to be consistent: in any state, it should evaluate logically-equivalent events in the same way.
}
Intuitively, evaluating a property with an oracle works as follows.
An execution sequence is processed in a lock-step manner.
On each received event, the oracle produces an appraisal on the sequence read so far.
For the formal semantics of the oracle and a formal definition of sequence checking, we refer to~\cite{DBLP:conf/rv/FalconeFM09}.
Figure~\ref{fig:runtime-oracle} shows an example of a runtime oracle that observes $e_1^* \cdot e_2^*$, where $e_1$ (resp. $e_2$) denotes that the variable $x$ in component $comp_1$ is strictly positive (resp. strictly negative).
\paragraph{Enforcement Monitor.}
An enforcement monitor (EM) is a finite-state machine that transforms a sequence of events from the program to one that evaluates on ``good verdicts" of the oracle.
The remaining description of the EM and how it interacts with the system serves as an abstract description of our instrumentation of CBS in \secref{sec:re-bip}.
Compared to enforcement monitors proposed in the literature, the ones introduced in this paper feature the ability to emit \emph{cancellation events} to revert the system back to a state where the underlying property is satisfied.
\begin{definition}[Enforcement monitor]
\label{def:em}
The enforcement monitor associated to the runtime oracle $\cO=\tuple{\Theta^\cO,\tinit\cO,\Sigma,\trans{}\cO,\bfour,\verdict^\cO}$ is a tuple ${\cE}=\tuple{\Theta^\cE , \tinit\cO , \Sigma\cup \overline{\Sigma},\trans{}\cE}$ where:
\begin{itemize}
\item
$\Theta^\cE \subseteq \Theta^\cO \cup \overline{\Theta}^\cO$ with
$\overline{\Theta}^\cO = \setof{\theta_e \mid e \in \Sigma \wedge \theta\in\Theta^\cO}$ s.t. $\Theta^\cE$ is reachable from $\tinit\cO$ with $\trans{}{\cE}$,
\item
$\overline{\Sigma} = \setof{\overline{e}\mid e \in \Sigma}$ is the set of \emph{cancellation events},
\item $\trans{}{\cE}$ is the transition function defined as
  $\trans{}{\cE} = \setof{\tuple{\theta,e,\theta'}\in\trans{}\cO \mid \verdict^\cO(\theta') \in \setof{\top,\top_c}}$\\
 $\cup \setof{ \tuple{\theta,e,\theta_e}, \tuple{\theta_e,\overline{e},\theta} \mid \exists \theta'\in \Theta: \tuple{\theta,e,\theta'} \in \trans{}\cO \wedge \verdict^\cO(\theta') = \bot}$.
\end{itemize}
\end{definition}
Intuitively, an enforcement monitor follows the structure of a runtime oracle on currently-good and good locations.
For each transition $\tuple{\theta,e,\theta'}$ leading to a ``bad" location $\theta'$ ($\verdict^\cO(\theta') = \bot$), the transition relation is modified in that we add a transition $\tuple{\theta, e ,\theta_e}$ leading to a fresh intermediate state $\theta_e$ and a transition $\tuple{\theta_e, \overline{e} ,\theta}$ back to the starting state $\theta$ labelled by the corresponding cancellation event.
Note, $\trans{}{\cE} $ is complete w.r.t. $\Sigma$.

We define the composition of a system with an enforcement monitor.
\begin{definition}[Composition of a system with an enforcement monitor]
\label{def:compositionem}
Given a system whose behavior can be formalized by an LTS $L=\tuple{\Sigma' , \mathrm{Loc}, \mathrm{Trans}}$ over the alphabet $\Sigma'$, with $\mathrm{Trans} \subseteq \mathrm{Loc} \times \Sigma \times \mathrm{Loc}$, and an enforcement monitor $\cE = \tuple{\Theta^\cO \cup \overline{\Theta}^\cO,\tinit\cO,\Sigma\cup \overline{\Sigma},\trans{}{\cE}}$ with $\Sigma \subseteq \Sigma'$ and $\overline{\Sigma} \cap \Sigma' = \emptyset$ for a safety property where states in $\Theta^\cO$ are associated to currently good and good verdicts, the composition is the LTS $\tuple{\mathrm{Loc} \times (\Theta^\cO \cup \overline{\Theta}^\cO), \Sigma\cup \overline{\Sigma}, \mathrm{Mon}}$, noted $L \otimes EM$, where the transition relation $\mathrm{Mon} \subseteq \mathrm{Loc} \times (\Theta^\cO \cup \overline{\Theta}^\cO) \times \Sigma \cup \overline{\Sigma} \times \mathrm{Loc} \times (\Theta^\cO \cup \overline{\Theta}^\cO)$ is defined by the two following semantics rules:
\[
\begin{array}{ccc}
\inferrule
{
q \trans{e'}{\mathrm{Trans}} q' \and e'\in\Sigma' \setminus \Sigma
} 
{
\tuple{q,\theta} \trans{e}{\mathrm{Mon}} \tuple{q',\theta}
} \, (1)
&
~\hspace{3em}~
&
\inferrule
{
\theta \trans{e}{\cE} \theta' \and \theta'\in \Theta^\cO \and q \trans{e}{\mathrm{Trans}} q' \and e\in\Sigma 
}
{
\tuple{q,\theta} \trans{e}{\mathrm{Mon}} \tuple{q',\theta'}
} \, (2)
\\
\multicolumn{3}{c}{
\inferrule
{
\theta \trans{e}{\cE} \theta_e \and \theta_e\in \overline{\Theta}^\cO \and \theta_e \trans{\overline{e}}{\cO} \theta \and q \trans{e}{\mathrm{Trans}} q'
}
{
\tuple{q,\theta} \trans{e\cdot \overline{e}}{\mathrm{Mon}} \tuple{q,\theta}
} \, (3)
}
\end{array}
\]
\end{definition}
At runtime, an enforcement monitor executes in a lock step manner with the system.
When the system emits an event that is not in the alphabet of interest of the enforcement monitor (i.e., an event $e' \in \Sigma' \setminus \Sigma$), the enforcement monitors lets the system execute without intervening (first semantics rule).
When the system emits an event that leads to a currently-good or good location, the enforcement monitor simply follows the system (second semantics rule).
When the system emits an event that leads to a bad location according to the oracle, the enforcement monitor executes a cancellation event.
In the third semantics rule, state $q'$ is called an unstable state: it is a state where the system never actually stays in because the enforcement monitor inserts immediately a cancellation event.
During an execution inserting the event $\overline{e}$ ``reverts" the effect of the event $e$ on the system: after an execution sequence $\sigma\in\Sigma^\ast$, for any event $e\in\Sigma$ and its associated cancellation event $\overline{e} \in \overline{\Sigma}$, the sequence $\sigma \cdot e \cdot \overline{e}$ ``amounts" to the sequence $\sigma$.
More formally, we define a function $\controlled$ between sequences of $(\Sigma\cup\overline{\Sigma})^\ast$ and sequences of $\Sigma^*$, inductively as follows:
\begin{itemize}
\item
$ \controlled(\epsilon) = \epsilon$,
\item
$\controlled(\sigma \cdot e) = \tau \cdot e$ if $\controlled(\sigma)= \tau$ and $e\in \Sigma$,
\item
$\controlled(\sigma \cdot e \cdot \overline{e}) = \tau$ if $\controlled(\sigma) = \tau$, $e\in \Sigma$, and $\overline{e} \in \overline{\Sigma}$.
\end{itemize}
Not all sequences in $(\Sigma \cup \overline{\Sigma})^\ast$ are in relation with a sequence in $\Sigma^\ast$ but the traces of  an LTS composed with an enforcement monitor are.

Given a system emitting events over $\Sigma$ and a safety property over $ \Sigma$.
Consider the composition of the enforcement monitor (obtained from the property) and the system.
Any execution of the composition projected on $\Sigma'$ deviates from the property by at most 1 event before being corrected, as stated by the following proposition.
\begin{proposition}
\label{prop:correctness-abstract}
Given a safety property $\Pi \in\Safety(\Sigma)$, its enforcement monitor as per Definition~\ref{def:em} (built from the associated runtime oracle), and a system whose behavior can be modeled by an LTS $L$, we have:
\begin{enumerate}
\item
$
\forall \sigma \in \traces(L \otimes EM): \left( \controlled(\sigma) \notin \Pi \wedge \exists e \in \Sigma,\exists \tau \in \Sigma^\ast: \sigma = \tau \cdot e \right) \implies \tau \in \Pi
$,
\item
$
\forall \sigma \in \traces(L \otimes EM): \controlled(\sigma) \in \Pi \cap \traces(L)
$
.
\end{enumerate}
\end{proposition}
Item $1$ states that the incorrect monitored traces that terminate with an event in $\Sigma$ (i.e., the traces that have not been corrected by enforcement monitors) have their longest maximal strict prefix correct w.r.t. $\Pi$.
Item $2$ states that the sequences associated to the monitored traces via function $\controlled$ are i) correct w.r.t. $\Pi$ and ii) belong to the original LTS.
\section{Runtime Enforcement for BIP Systems}
\label{sec:re-bip}
%
We instrument and integrate a runtime oracle ${\cal O} = \tuple{ \Theta^{\cal O}, \tinit{\cal O}, \Sigma,\trans{}{\cal O}, \bfour, \verdict^{\cal O} }$ for some (enforceable) property into a BIP system $\big(\pi(\Gamma(\setof{B_1,\ldots,B_n})),\tuple{l_0^1,\ldots,l_0^n}\big)$ made of a composite component $\pi(\Gamma(\setof{B_1,\ldots,B_n}))$, where the initial locations of the atomic components $B_1,\ldots, B_n$ are $l_0^1, \ldots, l_0^n$, respectively.
Some of the transformations proposed in this section are defined w.r.t. a particular component in the system.
For this purpose, we consider an atomic component $B = \tuple{P , L , T , X , \setof{g_\tau}_{\tau \in T},\setof{f_{\tau}}_{\tau \in T}}$ (cf. Definition~\ref{def:atomic}).
%
\subsection{Analysis and Extraction of the Needed Information}
\label{sect-enforcement-extraction}
%
The first step is to retrieve from the runtime oracle the set of components and their corresponding variables that should be monitored. 
For a property expressed over $\Sigma(\pi(\Gamma(\setof{B_1,\ldots,B_n})))$ and its oracle:
\begin{itemize}
\item
$\monvars(B_i)$ is the set of variables used in the property (that should be monitored) related to component $B_i$, formally
$\monvars(B_i) \defas \setof{ B_i.x \mid \exists e\in \Sigma, \exists \mathit{ap} \in \Prop(e): (B_i,x) \in \used({\it ap}) }$,
\item
$\occur$ is the set of all monitored variables, formally $\occur \defas \bigcup_{i\in [1,n]} \monvars(B_i)$.
\end{itemize}
For instance for the property described by the runtime oracle in \figref{fig:runtime-oracle}, we have $\monvars(\mathit{comp1}) = \setof{ \mathit{comp1}.x}$.
%
\subsection{Instrumenting Transitions}
%
To instrument the system in such a way that enforcement is as efficient as possible, we should only instrument the transitions that may modify some monitored variables. 
We denote by $\selecttrans(B)$ the set of the transitions that should be instrumented in $B$.
A transition is instrumented if either: (1) it modifies some monitored variables through its sequence of assignments; or (2) some monitored variables are assigned to its port. 
Note that, if the property contains a predicate on the location or on a port of a component $B$ (e.g., if $B.\dotloc = l_0$ appears in the property), then all transitions of that component should be instrumented.
Formally:
\[
  \selecttrans(B) \defas 
  \begin{cases}
    \transof{B} \quad \qquad \qquad \qquad \qquad \qquad \text{ if }  \setof{ B.\dotloc, B.\dotport} \cap \monvars(B) \neq \emptyset, \\
    \setof{\tau \in \transof{B} \mid \monvars(B) \cap (var(\tau.\func) \cup \varsof{\tau.\port}) \neq \emptyset} \quad  \text{ otherwise}.
  \end{cases}
\]
For the component in \figref{fig:atomic}, we have $\selecttrans(\mathit{comp1}) = \setof{(l,p, x>0, [y:= x+t],l')}$ since variable $x$ is both attached to port $p$ and $\mathit{comp1}.x \in \monvars(\mathit{comp1})$.

Instrumenting a transition consists in splitting it into four transitions as follows.
First, we reconstruct the initial transition.
Second, we create a transition to interact with the enforcement monitor through port $p^m$.
Finally, we create two transitions: one  to recover (through port $p^r$) when the property is violated and another to continue (through port $p^c$) otherwise.
In case of recovery, the modified variables are restored.
The ports $p^m, p^r, p^c$ are special, their purpose will be detailed in \secref{sec:re-bip:instrumenting}.
Formally, instrumenting a transition is defined by function $\insttrans$ that takes a transition and returns a set of four transitions as follows.
\begin{definition}[Instrumenting a transition]
\label{def:insttransition}
For any transition $\tau = \tuple{l,g,p,f,l'}$ in $T$, $\insttrans(\tau) = \setof{\tau^{i},\tau^{m},\tau^{c},\tau^{r}}$, where:
\begin{itemize}
  \item $\tau^{i} = \tuple{l,g,p,f^{i},l_m}$, where: 
{
\[
f^{i} =
  \begin{cases}
f &\text{if } B_i.\mathit{loc} \notin \monvars(B_i) \wedge B_i.\mathit{port} \notin \monvars(B_i),\\
f;[ \mathit{loc}:=\text{``$l'$"}] &\text{if } B_i.\mathit{loc} \in \monvars(B_i)  \wedge B_i.\mathit{port} \notin \monvars(B_i),\\
f;[\mathit{port}:= \text{``$p$"} ] &\text{if } B_i.\mathit{loc} \notin \monvars(B_i)  \wedge B_i.\mathit{port} \in \monvars(B_i),\\
f;[\mathit{loc}:=\text{``$l'$"};\mathit{port}:=\text{``$p$"}] &\text{if } B_i.\mathit{loc} \in \monvars(B_i) \wedge B_i.\mathit{port} \in  \monvars(B_i) ,
\end{cases}
\]
}
\item
$\tau^{m} = \tuple{l_m, \true, p^{m}, [~], l_r}$,
\item
$\tau^{c} = \tuple{l_r, \true, p^{c}, [~], l'}$,
\item $\tau^{r} = \tuple{l_r, \true, p^{r}, f^{r}, l}$, where 
  $f^{r} = [x_1 := x_1^{\tmp}; \ldots; x_j := x_j^{\tmp}]$ with $\setof{x_1,\ldots,x_j} = \setof{x \mid x \in \varsof{p} \vee x := f^x(X) \in f}$.
\end{itemize}
\end{definition}
\begin{example}[Instrumenting a transition]
Figure~\ref{fig:inst-atom} shows how the transition, in red in \figref{fig:atomic}, is instrumented.
On recovery, we restore all the variables that are modified when executing that transition.
Recall that some of the variables could be modified indirectly through the port of the transition ($p$), e.g., $x$ and $z$.
\end{example}
%
Recall that an interaction synchronizes a set of transitions and its execution implies firing all its corresponding transitions.
Hence, recovering implies to restore the previous global state of the system. 
For this purpose, instrumenting a transition $\tau \in \selecttrans(B_i)$ implies the instrumentation of all transitions synchronizing with $\tau$ through an interaction. 
We define $\rectrans$ to be the set of all transitions that should be instrumented. 
We also define $\reccomp$ to be the set of components that contain at least one instrumented transition, and  $\recinter$ to be the set of  connectors synchronizing on at least one instrumented transition.
Formally:
\[
\begin{array}{rcl}
  \rectransi &\defas& \cup_{i \in [1,n]} \selecttrans(B_i), \\
  \rectrans &\defas& \rectransi \cup \setof{\tau \mid \exists \gamma \in \Gamma, \exists \tau_k \in \rectransi: \setof{\tau.\port,\tau_k.\port} \subseteq P_{\gamma}}, \\
 \reccomp &\defas& \setof{B_i \mid B_i.T \cap \rectrans \neq \emptyset}, \\
  \recinter &\defas& \setof{a \in \Gamma \mid \exists \tau\in \rectrans:  \tau.\port \in P_{\gamma}}.
\end{array}
\]
\subsection{Instrumenting Atomic Components}
\label{sec:re-bip:instrumenting}
%
Let $T^{r}_B = \rectrans \cap \transof{B}$ be the set of transitions that should be instrumented in $B$ (noted $T^{r}$ when clear from context).
We create new temporary/recovery variables used to store the values of the variables that could be modified on an instrumented transition.
More precisely, for each variable that can be modified through a function or attached to a port of an instrumented transition, we create a corresponding temporary variable for it.
Given a set of transitions, we define the set of variables that should be recovered as follows: $\recvars(T') \defas \bigcup_{\tau \in T'} \varsof{\tau.\port}\, \cup\, var(\tau.\func)$.
If the enforcement monitor needs to observe the location or the port being executed, we create two new variables\footnote{Variables created by the transformations have fresh name w.r.t. existing variables of the input system.} $\mathit{port}$ and $\mathit{loc}$ that store the name of the next location and the name of the port being executed. 
We create three new ports:
\begin{enumerate}
\item
$p^m$ is used to \emph{send the value of the monitored variables} to the enforcement monitor;
\item
$p^c$ is used to receive a \emph{continue} notification from the enforcement monitor;
\item
$p^r$ is used to receive a \emph{recovery} notification from the enforcement monitor. 
\end{enumerate}
Finally, we split each of its instrumented transitions, that is $T^{r}$, according to Definition~\ref{def:insttransition}, and we create new locations accordingly.
Formally, instrumenting an atomic component is defined as follows:
\begin{definition}[Instrumenting an atomic component]
\label{def:instatomic}
We define the instrumentation function $\instf$ that transforms an input atomic component:
\[
\instf(B)=
  \begin{cases}
B  &\text{ if } B \notin \reccomp,\\
\tuple{P^{\instscript},L^{\instscript},T^{\instscript},X^{\instscript},\setof{g_\tau}_{\tau \in T^{\instscript}},\setof{f_{\tau}}_{\tau \in T^{\instscript}}} & \text{ otherwise.}
  \end{cases}
\]
where:
\begin{itemize}
\item
$X^{\instscript} = X \cup \setof{ v \mid \mathit{B_i.v} \in \monvars(B_i)} \cup \setof{x^{\tmp} \mid x \in \recvars(T^{r})}$ where, if $B_i.\dotloc \in \monvars(B_i)$ (resp. $B_i.\dotport \in \monvars(B_i)$), $\dotloc$ (resp.  $\dotport$) is initialized to $l_0^i$ (resp. $\mathtt{null}$), recovery/temporary variables are initialized to the values of their corresponding variables,
\item
  $P^{\instscript} = P \cup \setof{\tuple{p^{m},\monvars(B_i)},\tuple{p^{c},\emptyset},\tuple{p^{r},\emptyset}}$,
\item
$L^{\instscript} = L \cup \setof{l^m_{\tau} \mid \tau \in T^{r}}  \cup \setof{l^r_{\tau} \mid \tau \in T^{r}}$,
\item $T^{\instscript} = (T \setminus T^{r}) \ \cup \ (\bigcup_{\tau \in T^{r}} \insttrans(\tau))$.
\end{itemize}
\end{definition}
\begin{figure}[h]
\centering
\scalebox{0.7}{
\ifthenelse{\boolean{pdf}}
{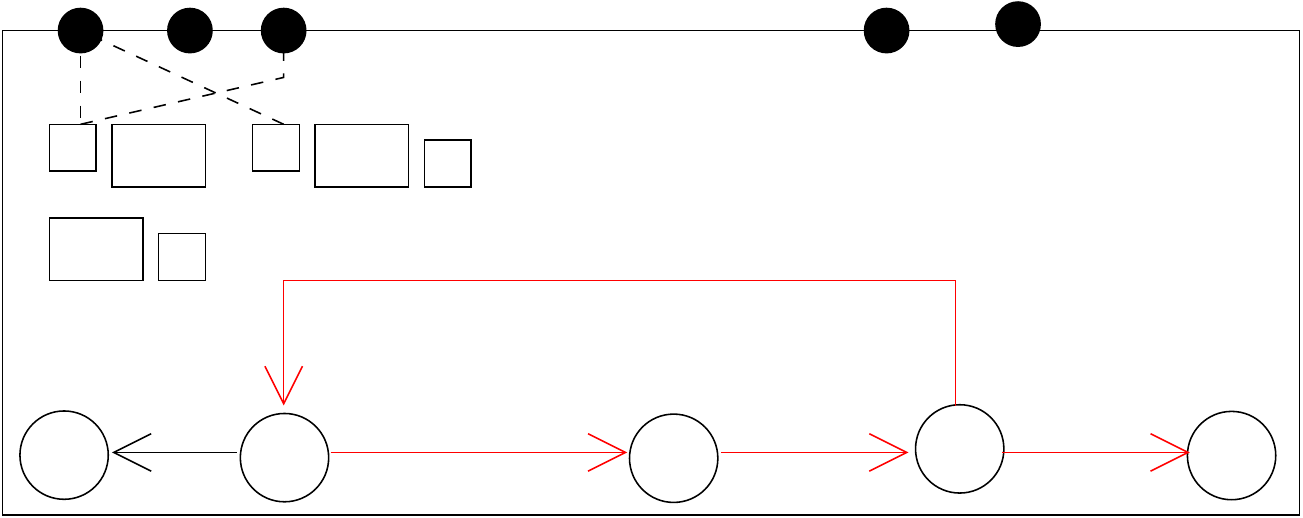}
{\input{fig/inst_atomic2.pstex_t}}
}
\caption{Instrumenting an atomic component}
\label{fig:inst-atom}
\end{figure}
\begin{example}[Instrumenting an atomic component]%
Figure~\ref{fig:inst-atom} shows the instrumentation of the atomic component in Fig.~\ref{fig:atomic}.
Note that only the transition in red is instrumented.
Also, the variables attached to port $p^m$ (i.e., only $\mathit{comp1}.x$ in this example) are those extracted from the oracle (see \figref{fig:runtime-oracle}), i.e., monitored variables of that component.
Moreover, the function of the recovery transition (i.e., labelled with with $p^r$) recovers the variables that could be modified, i.e., $x,y$, and $z$ since variables $x$ and $z$ are attached to port $p$ and $y$ is assigned on the transition.
\end{example}
In the sequel, we consider an instrumented atomic component $B^{\instscript} = \instf(B)$.
After instrumenting an atomic component, we must also create a backup of the variables that could be modified after executing an instrumented transition. 
That is, we need to store the values of those variables in their corresponding temporary variables.
For each transition, we select all the transitions of the next state that
are instrumented, and we backup the variables that could be modified on them.
\begin{definition}[Backup injection]
\label{def:backupInjection}
The backup injection function $\inj$ applied to $B^{\instscript}$ is the composite component $\inj(B^{\instscript}) = B^{\rec} = \tuple{ P^{\instscript},L^{\instscript},T^{\rec}, X^{\instscript},\setof{g_\tau}_{\tau \in T^{\rec}},\setof{f_{\tau}}_{\tau \in T^{\rec}} }$, where:
\[
\begin{array}{l}
T^{\rec} = \Big\{\tuple{l,g,p,f;[x_1^{\tmp} :=
  x_1; \ldots; x_j^{\tmp} := x_j],l'}\\  \qquad \mid  \tau = \tuple{l,g,p,f,l'} \in
T^{\instscript} \wedge \setof{x_1,\ldots,x_j} = \recvars(\setof{\tau^i \in B^{\instscript}.T^{r} \mid  \tau^i.\source = l'  \wedge \tau^i.\port \in P}
)\Big\}.
\end{array}
\]
\end{definition}
In the sequel, we consider an atomic component with injected backup $B^{\rec} = \inj(B^{\instscript})$.
\begin{figure}[t]
\centering
\scalebox{0.9}{
\ifthenelse{\boolean{pdf}}
{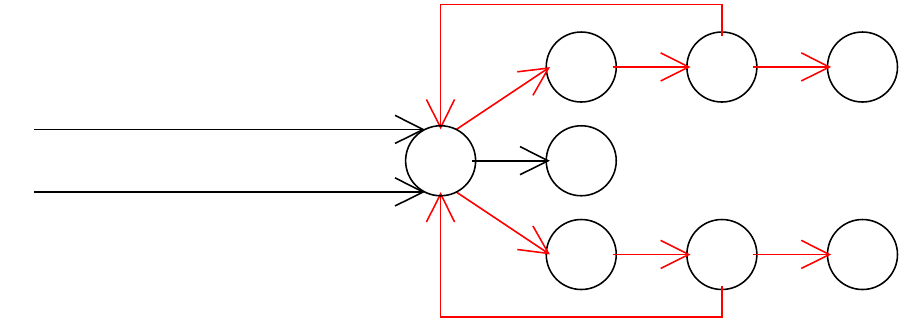}
{\input{fig/inst_injection.pstex_t}}
}
\caption{Injecting backup into an atomic component.}
\label{fig:inst-atomic}
\end{figure}
\begin{example}[Backup injection]
Figure~\ref{fig:inst-atomic} shows an example of backup injection (depicted in blue), into an instrumented atomic component.
Variables $x$ and $z$ are backed up on any transition entering $l_0$ because there are two outgoing transitions from $l_0$ that modify variables $x$ and $z$.
\end{example}
%
\subsection{Creating a BIP Enforcement Monitor from an Oracle}
\label{sect-recbs-creatingatomic}
%
We present how a runtime oracle ${\cal O}$ is transformed into a BIP enforcement monitor ${\cal E}$  that mimics the behavior of the enforcement monitor associate to ${\cal O}$ (see Definitions \ref{def:oracle} and \ref{def:em}). The generated BIP enforcement monitor receives events from the instrumented atomic components and processes them to produce the same verdicts as the initial abstract oracle.
Depending on the state of ${\cal E}$, it notifies the instrumented atomic components to continue or to recover.  
\begin{figure}[b]
\centering
\scalebox{0.5}{
\ifthenelse{\boolean{pdf}}
{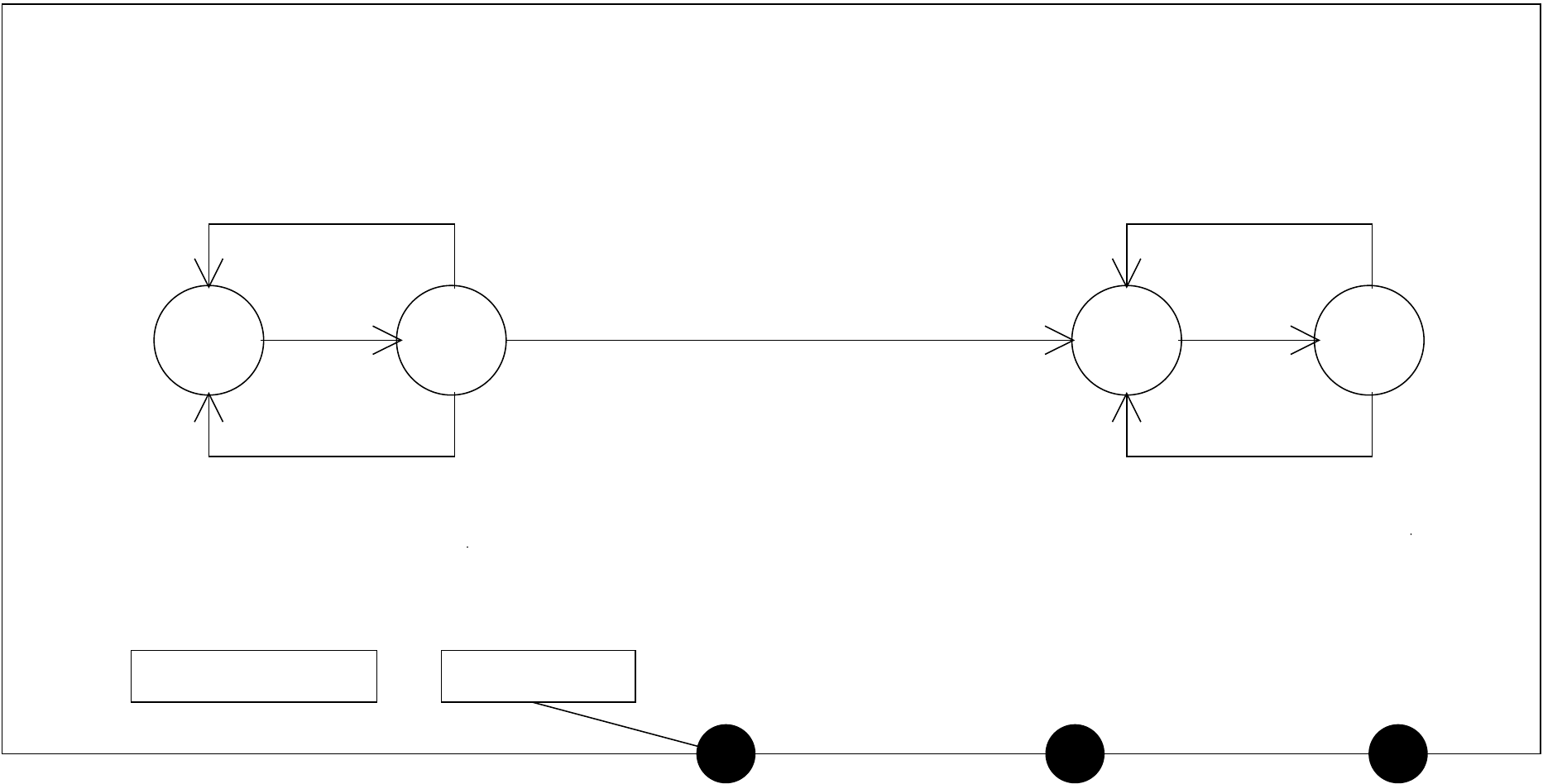}
{\input{fig/monitor2.pstex_t}}
}
\caption{Enforcement monitor}
\label{fig:monitorexample}
\end{figure}
The enforcement monitor contains a copy of the monitored variables and a backup/temporary copy of them. 
When the instrumented system executes an interaction that synchronizes at least one instrumented transition, it interacts with the enforcement monitor through port $p^m$ and sends the modified values of monitored variables.
Depending on those values the enforcement monitor produces a verdict and notifies the original system to continue or to recover, accordingly.
In case of recovery (resp. continue), the supervised system should also recover (resp. backup) its monitored variables. The behavior of the enforcement monitor is formalized as follows. 
\begin{definition}[Building an enforcement monitor]
\label{def:generatemonitor}
From oracle ${\cal O}$ we define the enforcement monitor ${\cal E}=\tuple{P,L,T,X,\setof{g_\tau}_{\tau \in T}, \setof{f_{\tau}}_{\tau \in T}}$ as an atomic component:
\begin{itemize}
\item $X= \occur \cup X^{\tmp}$ with $X^{\tmp} = \setof{x^{\tmp} \mid x \in \occur}$,
\item $P = \setof{\tuple{p^{m},\occur},\tuple{p^{c},\emptyset},\tuple{p^{r},\emptyset}}$,
\item $L = L^{\top} \cup L^m$ with $L^{\top} = \setof{q \mid q \in \Theta^{\cal O} \wedge \verdict^{\cal O}(q) \in \setof{\top, \top_c}}$ and $L^m = \setof{q^m \mid q \in L^{\top}}$, 
\item $T = T^{m} \cup T^{r} \cup T^{c}$ with
\begin{itemize}
\item
$T^m = \setof{\tuple{q,p^m,\true,[~],q^{m}}\mid q \in L^{\top}}$;
\item
$T^c=\setof{\tuple{q^m,p^{c},e,f^c,q'} \mid q \trans{e}{{\cal O}} q' \wedge \verdict^{\cal O}(q') = \top}$, where $f^{c} = [x_1^{\tmp} := x_1; \ldots; x_j^{\tmp} = x_j]$ with $\tuple{x_1^{\tmp},\ldots,x_j^{\tmp}} = X^{\tmp}$;
\item
$T^r=\setof{\tuple{q^m,p^{r},e,f^r,q} \mid q \trans{e}{{\cal O}} q' \wedge \verdict^{\cal O}(q') = \bot}$, where $f^{r} = [x_1 := x_1^{\tmp}; \ldots; x_j := x_j^{\tmp}]$ with $\tuple{x_1^{\tmp},\ldots,x_j^{\tmp}} = X^{\tmp}$.
\end{itemize}
\end{itemize}
\end{definition}
\begin{example}[Building an enforcement monitor]
Figure~\ref{fig:monitorexample} depicts the enforcement monitor in BIP generated from the runtime oracle in \figref{fig:runtime-oracle}.
From the initial state, the enforcement monitor synchronizes with the system by receiving the value of $\mathit{comp_1.x}$ through port $p^m$.
Then, depending on the value of $\mathit{comp_1.x}$, it either recovers (when $\mathit{comp_1.x}$ is equal to $0$, or continues otherwise.
In case of continue, variable $\mathit{comp_1}.x$ is backed up.
In case of recovery, variable $\mathit{comp_1}.x$ is recovered.
\end{example}
%
\subsection{Integration - Spin Recovery}
\label{sec:integration:spin}
%
We define the connection between the instrumented atomic components $\pi(\Gamma(\setof{B_1^{\rec},\ldots,B_n^{\rec}}))$ and enforcement monitor ${\cal E}$.
We connect the $p^m$ ports of the instrumented components with the $p^m$ port of ${\cal E}$ ($\gamma_{m}$).
All the ports of that connector should be trigger to make all interactions possible. 
Because of maximal progress, all the enabled $p^m$ ports of the instrumented components will be synchronized with the port $p^m$ of ${\cal E}$. 
The update function of that connector transfers the updated values of the monitored variables from the instrumented atomic components to ${\cal E}$. 

Then, we connect all the continue ports of the instrumented atomic components, i.e. $p^c$,  with a connector where its ports are marked as trigger.
This connector will be connected hierarchically to the port $p^c$ of ${\cal E}$. The ports of the hierarchical connector are marked  as synchron so that the synchronization between the $p^c$ port of the instrumented components requires the port $p^c$ of ${\cal E}$ to be enabled.
This is necessary because the instrumented components will be ready to execute both the continue and the recoverable ports based on the decision taken by ${\cal E}$.
In the same way, we connect the recoverable ports. 

Finally, the priority model is augmented by giving more priority to the interactions defined by the monitored, continue, and recoverable connections.
Modifying the priority model ensures that, after the execution of an interaction synchronizing some instrumented transition, ${\cal E}$ notifies the system to recover or to continue before involving other interactions synchronizing instrumented transitions.

Note that, when some of the ports $p^m$ of the instrumented atomic components are enabled, the port $p^m$ of ${\cal E}$ is also enabled.
However, the instrumented atomic components could be in a state where none of their $p^m$ ports are enabled.
To prevent ${\cal E}$ from moving without synchronizing with the components, the port $p^m$ of ${\cal E}$ is synchron. 
\begin{definition}[Integration - spin recovery]
\label{def:connectionsspin}
The composite component is $\pi^{\rec}(\Gamma^{\rec}(B_1^{\rec},\ldots,B_n^{\rec},{\cal E}))$, where:
\begin{itemize}
\item $\Gamma^{\rec} = \Gamma \cup \setof{\gamma^{m},\gamma^{c_1},\gamma^{c_2},\gamma^{r_1},\gamma^{r_2}}$, where:
\begin{itemize}
  \item $\gamma^{m} = \tuple{P_{\gamma^{m}},t_{\gamma^{m}},\true,F_{\gamma^{m}}}$, where:
\begin{itemize}
  \item $P_{\gamma^{m}} = \setof{\tuple{B_i.p^m, \monvars(B_i)}}_{B_i \in \reccomp} \cup \setof{{\cal E}.p^m}$, $t_{\gamma^m}({\cal E}.p^m) = \false$ and 
$\forall p \in P_{\gamma^{m}} \setminus \setof{{\cal E}.p^m}: t_{\gamma^m}(p) = \true$,
\item $F_{\gamma^{m}}$, the update function, is the identity data transfer from the variables in the ports of the interacting components to the corresponding variables in the oracle port.

\end{itemize} 
\item $\gamma^{c_1} = \tuple{P_{\gamma^{c_1}},t_{\gamma^{c_1}},\true,[~]}$, $\gamma^{c_2} = \tuple{P_{\gamma^{c_2}},t_{\gamma^{c_2}},\true,[~]}$, where:
\begin{itemize}
  \item $P_{\gamma^{c_1}} = \setof{\tuple{B_i.p^c,\emptyset}}_{B_i \in \reccomp}$ and $\forall p \in P_{\gamma^{c_1}}: t_{\gamma^{c_1}}(p) = \true$,
\item $P_{\gamma^{c_2}} =  \setof{\gamma^{c_1}.\export, {\cal E}.p^c}$ and $t_{\gamma^{c_2}}(\gamma^{c_1}.\export) = t_{\gamma^{c_2}}({\cal E}.p^c) = \false$.
\end{itemize} 
\item $\gamma^{r_1} = \tuple{P_{\gamma^{r_1}},t_{\gamma^{r_1}},\true,[~]},\gamma^{r_2} = \tuple{P_{\gamma^{r_2}},t_{\gamma^{r_2}},\true,[~]}$, where:
\begin{itemize}
  \item $P_{\gamma^{r_1}} = \setof{\tuple{B_i.p^r, \emptyset}}_{B_i \in \reccomp}$ and 
$\forall p \in P_{\gamma^{r_1}}:t_{\gamma^{r_1}}(B_i.p^r) = \true$,
\item $P_{\gamma^{r_2}} =  \setof{\gamma^{r_1}.\export, {\cal E}.p^r}$ and $t_{\gamma^{r_2}}(\gamma^{r_1}.\export) = t_{\gamma^{r_2}}({\cal E}.p^r) = \false$,
\end{itemize} \end{itemize}
\item $\pi^{\rec}=\pi \cup \setof{\tuple{a,a'}\mid a \in \cup_{\gamma \in \recinter} {\cal I}(\gamma) \wedge a' \in {\cal I}(\gamma^m,\gamma^{c_1},\gamma^{c_2},\gamma^{r_1},\gamma^{r_2})}$.
\end{itemize}
\end{definition}
An example of integration with spin recovery is provided in the following sub-section.
\begin{remark}
If the system reaches a state, where no further transition is possible, the supervised system will enter in a livelock as all transitions will be tried and rolled back indefinitely.
\end{remark}
%
\subsection{Integration - With Disabler}
\label{sec:integration:disabler}
%
The instrumented system defined above may be inefficient in some cases.
For instance, when ${\cal E}$ notifies the system to recover, the system may execute  again one of the previously executed bad interactions.
To solve this issue, we create a disabler component that comes as an optimization for the monitored system.
The idea is to keep disabled the bad interactions that we have recovered from, until a good interaction is found (note: the system should contain at least one possible good interaction, which can possibly be taken after recovering, if no good interaction exists then the system would reach a deadlock state after the system has exhausted all available interactions).
For this purpose, we assume that all connectors of the input BIP system contain only synchron ports, hence each connector represents only one interaction.
In the following, we use the terms connector and interaction interchangeably.  

\begin{figure}[t]
\centering
\scalebox{0.5}{
\ifthenelse{\boolean{pdf}}
{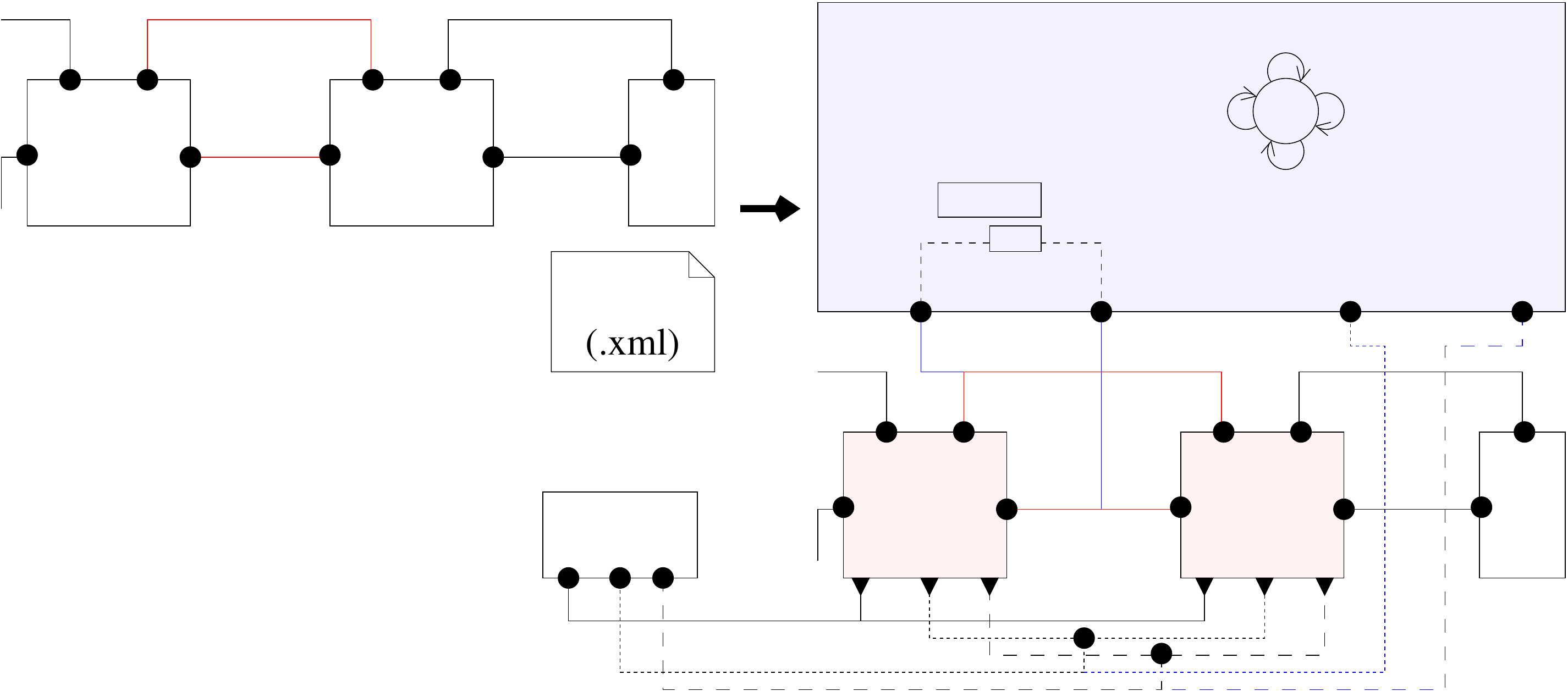}
{\input{fig/inst_compound.pstex_t}}
}
\caption{Supervised system [with/without] disabler (spin recovery -- without disabler).}
\label{fig:supervisedsystem}
\end{figure}

For each interaction ($a \in \recinter$) connected to an instrumented transition, we associate a transition in the disabler. This transition will be labeled with a port connected to the interaction that corresponds to that transition. That is, to execute that interaction, the port of the corresponding transition of the disabler should be ready as well.
On executing that interaction the id representing the interaction is sent to the disabler.
We also create a continue port $p^c$ and a recoverable port $p^r$ that will be synchronized with ${\cal E}$ in case of continue and recovery, respectively. 
The disabler synchronizes with ${\cal E}$ on the recovery and continue ports.
On a recovery, ${\cal E}$ synchronizes with the instrumented components and with the disabler.
The disabler will set the guard of the corresponding last received id to $\false$.
Consequently, after recovery, the last executed interaction cannot not be taken again. 
On continue, ${\cal E}$ informs the disabler that it should enable all its ports, by re-setting their corresponding guards to true, and now all interactions become valid.
For each recoverable interaction, i.e., $a \in \recinter$ we assign a positive integer for it: $\indexf: \recinter \rightarrow [0,|\recinter| - 1]$.
\begin{definition}[Disabler construction in BIP]
Given the set of recoverable interactions $\recinter$ we construct the disabler ${\cal D} = \tuple{P,L,T,X,\setof{g_\tau}_{\tau \in T},\setof{f_{\tau}}_{\tau \in T}}$, where: 
\begin{itemize}
  \item
  $P = \setof{p^r} \cup \setof{p^c} \cup \setof{\tuple{p^\gamma,\emptyset} \mid \gamma \in \recinter}$,
  \item
   $L = \setof{l}$,
	\item
	$X = \setof{\enab,id}$, where $\enab$ is an array of Booleans initialized to $\true$ and its size is equal to $|\recinter|$,
\item $T = T^r \cup T^c \cup T^{\inter}$, where:
\begin{itemize}
\item $T^r = \setof{\tau^r}$, where $\tau^r = \tuple{l,\true, p^r, \big[\enab[id] := \false\big],l}$,
\item $T^c = \setof{\tau^c}$, where $\tau^c = \tuple{l,\true, p^c, \big[\enab[0] := \true; \ldots; \enab[|\recinter| - 1] := \true \big],l}$,
\item $T^{\inter} = \setof{\tuple{l,\enab[\indexf(\gamma)],p^\gamma,[id := \indexf(\gamma)],l} \mid \gamma \in \recinter}$.
\end{itemize}
\end{itemize}
\end{definition}
\begin{example}[Disabler]
Figure~\ref{fig:supervisedsystem} provides an example of disabler in blue.
We have $\recinter = \setof{a_0, a_1}$ ($a_0$ and $a_1$ contain ports that are attached to instrumented transitions). The disabler contains transitions that correspond to $a_0$ and $a_1$. 
Those transitions are labeled with ports $p^{a_0}, p^{a_1}$ which are connected to interactions $a_0$ and $a_1$. Moreover, the disabler contains an array of Boolean variables of size 2. The transitions that correspond to $\recinter$ are guarded with the elements of the array accordingly. 
In case of recovery, e.g., after executing $a_0$ (resp. $a_1$), the corresponding Boolean variable is set to $\false$, and hence, interaction $a_0$ (resp. $a_1$) is disabled. In case of continue, all the elements of the array are set to $\true$. 
\end{example}
As in Definition~\ref{def:connectionsspin}, we connect the instrumented system with ${\cal E}$, but we also connect the instrumented interactions to their corresponding ports of the disabler.
Moreover, we connect the continue port (resp. the recovery port) of ${\cal E}$ with the continue port (resp. the recovery port) of the disabler.
As in Definition~\ref{def:connectionsspin}, we augment the priority model. 
\begin{definition}[Integration - with disabler]
\label{def:connectionsdisabler}
Given an enforcement monitor in BIP ${\cal E}$ and a composite component $\pi(\Gamma(\setof{B_1^{\rec},\ldots,B_n^{\rec}}))$ obtained as in Definition~\ref{def:connectionsspin}, that is, $B_i^{\rec} = \inj(\instf(B_1^m))$, and disabler ${\cal D}$, we build the composite component $\pi^{\rec}(\Gamma^{\rec}(B_1^{\rec},\ldots,B_n^{\rec}, {\cal E}, {\cal D}))$, where,
\begin{itemize}
\item $\Gamma^{\rec} = (\Gamma \setminus \recinter) \cup \Gamma^{\recinter} \cup \setof{\gamma^m,\gamma^{c_1}, \gamma^{c_2}, \gamma^{r_1}, \gamma^{r_2}}$, 
\begin{itemize}
\item $\Gamma^{\recinter} = \setof{\gamma^{\recinter} = (P_{\gamma^{\recinter}}, t_{\gamma^{\recinter}},G_{\gamma}, F_{\gamma})}_{\gamma = (P_{\gamma},t_{\gamma},G_{\gamma},F_{\gamma}) \in \recinter}$ where
 $P_{\gamma^{\recinter}} = P_{\gamma} \cup \setof{p^\gamma \mid p^{\gamma} \in \gamma \in \recinter}$ and $\forall p \in P_{\gamma^{\recinter}}:t_{\gamma^{\recinter}}(p) = \false$;
\item $\gamma^m = (P_{\gamma^m},t_{\gamma^m},\true,F_{\gamma^m})$ , with
\begin{itemize}
  \item $P_{\gamma^m} = \setof{\tuple{B_i.p^m,\monvars(B_i)}}_{B_i \in \reccomp} \cup \setof{{\cal E}.p^m}$, 
  \item $t_{\gamma^m}({\cal E}.p^m) = \false$, and $\forall p \in P_{\gamma^m} \setminus \setof{{\cal E}.p^m}: t_{\gamma^m}(p) = \true$;
\item $F_{\gamma^m}$, the update function, is the identity data transfer from the variables in the ports of the interacting components $B_i$ ($i\in [1,n]$) to the corresponding variables in the oracle port; 
\end{itemize} 
\item $\gamma^{c_1} = (P_{\gamma^{c_1}},t_{\gamma^{c_1}},\true,[~])$, $\gamma^{c_2} = (P_{\gamma^{c_2}},t_{\gamma^{c_2}},\true,[~])$, with
\begin{itemize}
  \item $P_{\gamma^{c_1}} = \setof{\tuple{B_i.p^c,\emptyset}}_{B_i \in \reccomp}$ and $\forall p \in P_{\gamma^{c_1}}: t_{\gamma^{c_1}}(p) = \true$;
\item $P_{\gamma^{c_2}} =  \setof{\gamma^{c_1}.\export, {\cal E}.p^c, {\cal D}.p^c}$ and  
$\forall p \in P_{\gamma^{c_1}}: t_{\gamma^{c_2}}(p) = \false$;
\end{itemize} 
\item $\gamma^{r_1} = (P_{\gamma^{r_1}},t_{\gamma^{r_1}},\true,[~])$, $\gamma^{r_2} = (P_{\gamma^{r_2}},t_{\gamma^{r_2}},\true,[~])$, with
\begin{itemize}
  \item $P_{\gamma^{r_1}} = \setof{\tuple{B_i.p^r,\emptyset}}_{B_i \in \reccomp}$ and 
$\forall p \in P_{\gamma^{r_1}}: t_{\gamma^{r_1}}(p) = \true$;
\item $P_{\gamma^{r_2}} =  \setof{\gamma^{r_1}.\export,{\cal E}.p^r,{\cal D}.p^r}$ and  $\forall p \in P_{\gamma^{r_2}}: t_{\gamma^{r_2}}(p) = \false$;
\end{itemize} 
\end{itemize}
\item $\pi^{\rec} = \pi \cup \setof{\tuple{a,a'} \mid a \in \cup_{\gamma \in \recinter} {\cal I}(\gamma) \wedge a' \in {\cal I}(\gamma^m) \cup {\cal I}(\gamma^{c_1}) \cup {\cal I}(\gamma^{c_2}) \cup {\cal I}(\gamma^{r_1}) \cup {\cal I}(\gamma^{r_2})}$.
\end{itemize}
\end{definition}
\begin{example}[Integration - With Disabler]
Figure~\ref{fig:supervisedsystem} shows the supervised system with ${\cal E}$ and ${\cal D}$.
In case of spin recovery, we do not include ${\cal D}$ and its connections.
In this example, we assume that the monitored variables are modified only when executing interactions $a_0$ and $a_1$.
Consequently, component $B_3$ remains unchanged.
Notice that the expressiveness and modularity of BIP design allows us to add and remove ${\cal D}$ without modifying the behaviors of components.
\end{example}
\begin{remark}
If the system reaches a state, where no further transition is possible, it will enter in a deadlock as all transitions will be tried,  rolled back, and disabled successively by the disabler.
\end{remark}
%
\subsection{On the Correctness and Behavior of the Supervised System}
%
Correctly observing the system behavior relies on our instrumentation technique and follows the same correctness arguments as in~\cite{FalconeJNBB13}.
Correctness of the whole approach stems from the facts that we consider safety properties and that, as it was similarly expressed at an abstract level in Proposition~\ref{prop:correctness-abstract}, our enforcement monitors roll-back the system by one step as soon as the system emits an event that violates the property.

Intuitively, the correctness proof of the transformations consists in showing that the supervised BIP system behaves in the same way as the composition of an abstract enforcement monitor with the LTS of the initial system. That is, the behavior of the supervised systems follows the semantics rules in Definition ~\ref{def:compositionem}.
\paragraph{Preliminaries.}
A run of length $l$ of a system $(B,\mathit{Init})$ whose runtime semantics is $\pi(C)=(Q,A,\goesto_\pi)$ is the sequence of alternating states/configurations and interactions $q^0\cdot a_0\cdot q^1\cdot a_1\cdots a_{l-1}\cdot q^l$ such that: $q^0$ = $\mathit{Init}$, and, $\forall i\in [0,l-1]: q^i\in Q \wedge \exists a_i\in A: q^i \goesto[a_i]_\pi q^{i+1}$. 

Following the transformations defined in \secref{sec:re-bip}, a run $r = q^0\cdot a_0\cdot q^1\cdot a_1\cdots a_{l-1}\cdot q^l$ of the monitored system $C^{rec}$ satisfies the following properties: 
\begin{itemize}
\item if $\cE.p^m \in a^i$, then all other ports involved in $a^i$ are $p^m$ ports. The same property applies to $\cE.p^c$ and $\cE.p^r$. We denote by $\alpha_m$ (resp. $\alpha_c$, $\alpha_r$) any interactions involving $\cE.p^m$ (resp. $\cE.p^c$, $\cE.p^r$). This holds by construction according to Definitions~\ref{def:connectionsspin} and \ref{def:connectionsdisabler}.
\item Let $i \in [1, m]$ s.t. $q^i \cdot a^i \cdot q^{i+1}$, then $\cE.p^m \in a^i$ iff  $q^{i+1} \cdot a^{i+1} \cdot q^{i+2}$ where $\{\cE.p^r, \cE.p^c\} \cap a^{i+1} \neq \emptyset$. This stems from the following facts: (1) according to Definitions~\ref{def:connectionsspin} and \ref{def:connectionsdisabler}, $\alpha^c$ and $\alpha^r$ interactions have more priority than the interactions of the initial BIP system, and (2) according to Definition~\ref{def:insttransition}, an instrumented transition of an atomic component consists of a recovery and continue transitions just after a transition for interacting with the monitor (i.e., labeled with port $p^m$). 
\end{itemize}
Given a run $q^0\cdot a_0\cdot q^1\cdot a_1\cdots a_{l-1}\cdot q^l$ of the supervised system. Let us consider the next step of the system which consists in performing an interaction $a$. We distinguish two cases according to whether $a$ is connected to an instrumented transition (i.e., $a \in \recinter$) or not: 
\begin{enumerate}
\item If $a \notin \recinter$, then the execution of $a$ does not modify any variable of the property. This stems from the fact that: (1) according to Definitions~\ref{def:connectionsspin} and \ref{def:connectionsdisabler}, $\alpha^m$ interaction has more priority than the interactions of the initial BIP system, and (2) according to Definition~\ref{def:insttransition}, an instrumented transition of an atomic component consists of its previous transition followed by a transition to interaction with the monitor (i.e., labeled with port $p^m$).  This is mapped to an event $e' \in \Sigma' \setminus \Sigma$ in correspond to rule number 1 in Definition~\ref{def-runtimesemanticscomposite}.
\item If $a \in \recinter$, then $a$ is followed by the execution of a $\alpha^m$ interaction (i.e., interacting with the enforcement monitor). The values of the variables sent through the port $p^m$ of the enforcement monitor (i.e. ${\cal E}.p^m$) is mapped to a event $e \in \Sigma$ in Definition~\ref{def-runtimesemanticscomposite}. In that case, we distinguish two sub-cases:
\begin{enumerate}
\item When $a$ involves transitions that do not modify the variables of the property but at least one of these transitions has a port in an interaction that modifies certain variables of the property.
Henceforth, $e$ corresponds to the last emitted event in the run. Because of stutter invariance, the system keeps satisfying the property. This situation corresponds to rule number 2 in Definition~\ref{def-runtimesemanticscomposite}.
\item When $a$ involves transitions that modify some variables of the property, we distinguish two more sub-cases.
\begin{enumerate}
\item
When $e$ brings the monitor to a good (with verdict $\top$) or currently good state (with verdict $\top_c$), the system execute an $\alpha^c$ interaction that moves the system to a next good state (the same as in the original system).
This situation also corresponds to rule number 2 in Definition~\ref{def-runtimesemanticscomposite}.
\item
When $e$ brings the monitor to a bad state (with verdict $\bot$), the system execute an $\alpha^r$ interaction that restores the values of the variables and brings the system to its previous state which was correct. The execution of recovery corresponds to $\overline{e}$ in rule number 3 in Definition~\ref{def-runtimesemanticscomposite}.
\end{enumerate}
\end{enumerate}
\end{enumerate}
Notice that, if added, the disabler ${\cal D}$ might disable an interaction that violates the property and the scheduler would select the next one in terms of priority. For example, consider a composite component with two interactions $a_0$ and $a_1$ such that $a_0$ has more priority than $a_1$. 
If $a_0$ is always enabled, then according to the BIP semantics $a_1$ could not be enabled.
However, in the supervised system, if $a_0$ leads to a bad state, ${\cal D}$ will disable that interaction.
Consequently, interaction $a_1$ becomes enabled.
This can be seen as a powerful primitive to enforce the correctness of a system by allowing low priority interactions. 
However, in some cases, a property should be enforced while preserving the priority model.
In that case, on recovery, ${\cal D}$ must disable all interactions with less priority than the last executed one.

\subsection{Summary}
%
From the abstract oracle we generate the corresponding enforcement monitor in BIP. 
The instrumented system interacts with the enforcement monitor to avoid bad behaviors. 
The main idea of our method is to recover the system when the enforcement monitor detects a bad state. The recovering process is done as follows: 
(1) If the execution of an interaction modifies some monitored variables, the system should send the updated variable values to the enforcement monitor; 
(2) Depending on those values, the enforcement monitor interacts with the system and lets it continue or makes it recover the last correct state. 
After recovery, the system may take again the last executed interaction which will lead again to a bad state.
Such situation may lead the system to loop infinitely (livelock).
To prevent this, we define a \textit{disabler} component which is notified of a recovery to disable the bad interaction, until a good interaction is found.
%
%
%
\section{Implementation and Evaluation}
\label{sec:implem}
%
This section presents {\toolname} (see \secref{sec:implem:rebip}), an implementation of the transformations presented in \secref{sec:re-bip}, and its evaluation on two case studies: deadlock avoidance for dining philosphers (see \secref{sec:implem:deadlock}) and correct placement of robots (see \secref{sec:implem:robot})
%
\subsection{{\toolname}: a Toolset for Runtime Enforcement of BIP Systems}
\label{sec:implem:rebip}
%
\begin{figure}[t]
\centering
\ifthenelse{\boolean{pdf}}
{\includegraphics[width=\linewidth]{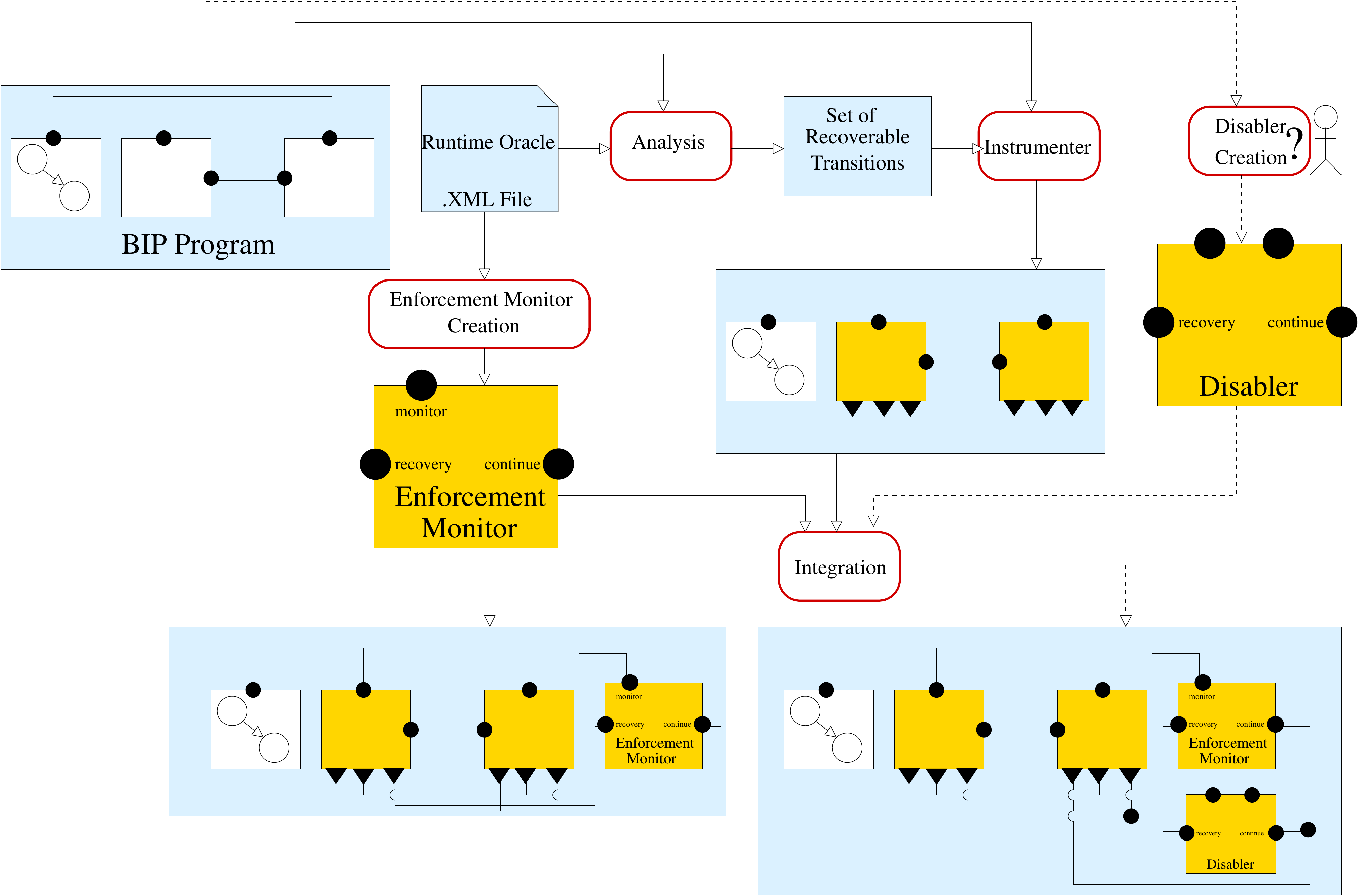}}
{\includegraphics[width=\linewidth]{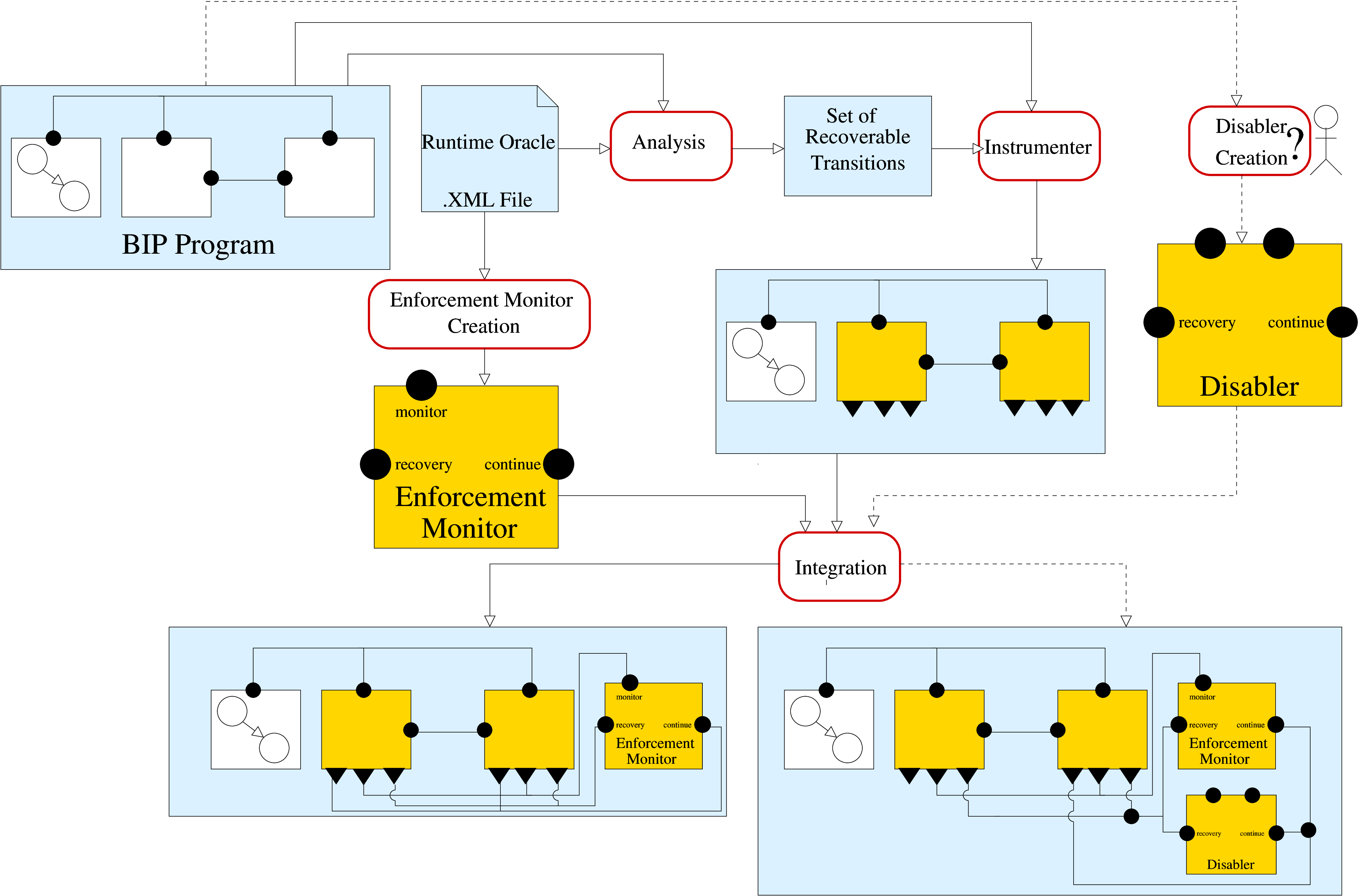}}
 \caption{Toolset for runtime enforcement ({\toolname})}
  \label{fig:toolre}
 \end{figure}
 {\toolname}\footnote{\url{http://ujf-aub.bitbucket.org/re-bip/}} is a Java implementation (8,000 LOC) of the transformations described in \secref{sec:re-bip}, and, is part of the BIP distribution. {\toolname} takes as input a BIP system and an abstract oracle (an XML file) and then outputs a new BIP system whose behavior is supervised at runtime (see \figref{fig:toolre}).
{\toolname} uses the following modules (see \figref{fig:toolre}):
\begin{itemize}
\item Analysis: from the runtime oracle of the property, collect the variables that should be monitored;
\item Instrumentation: according to the analysis, instrument some of the atomic components;
\item Enforcement Monitor Creation: from the runtime oracle (given as an XML file), generate the corresponding enforcement monitor in BIP;
\item Integration: according to the user's input, create the supervised system with or without the disabler. 
\end{itemize}
%
%
\subsection{Using {\toolname} to Avoid Deadlocks}
\label{sec:implem:deadlock}
\begin{figure}[t]
\centering
  \subfigure[Philosopher]{\label{fig:philo} \scalebox{0.55}{
\ifthenelse{\boolean{pdf}}
 {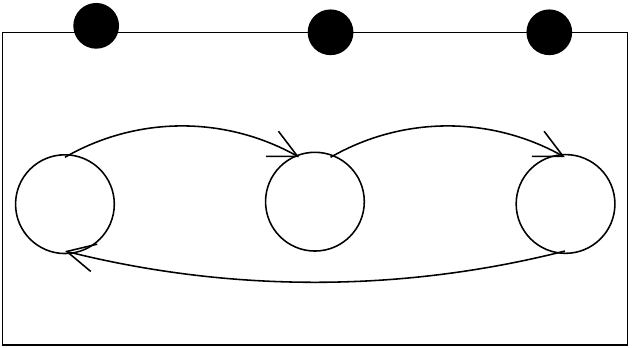}
 {\input{fig/philo.pstex_t}}
 }
 }
  \quad
   \subfigure[Fork]{\label{fig:fork} \scalebox{0.55}{
   \ifthenelse{\boolean{pdf}}
{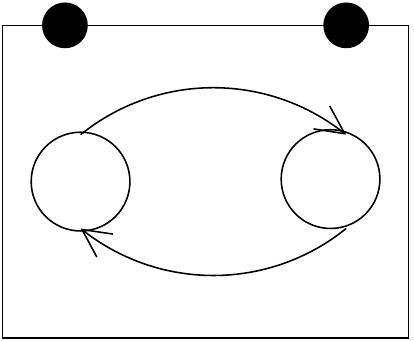}
{\input{fig/fork.pstex_t}}
}
} 
 \quad
  \subfigure[Dining philosophers in BIP]{\label{fig:dining} \scalebox{0.7}{
  \ifthenelse{\boolean{pdf}}
{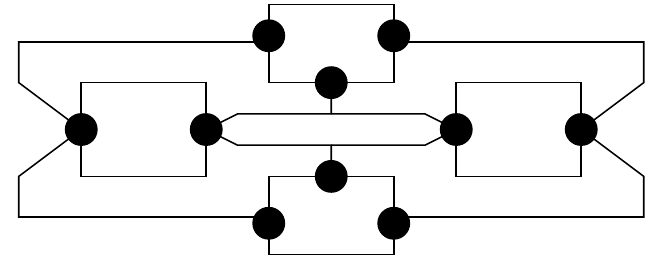}
{\input{fig/diningbip.pstex_t}}
}
}
 \caption{Dining philosophers with possible deadlock}
  \label{fig:diningbench}
\end{figure}

\begin{figure}[t]
\centering
  \subfigure[Comparison of execution times]{
  \label{bench:timedining}
  \scalebox{0.25}{
     \ifthenelse{\boolean{pdf}}
  {\includegraphics{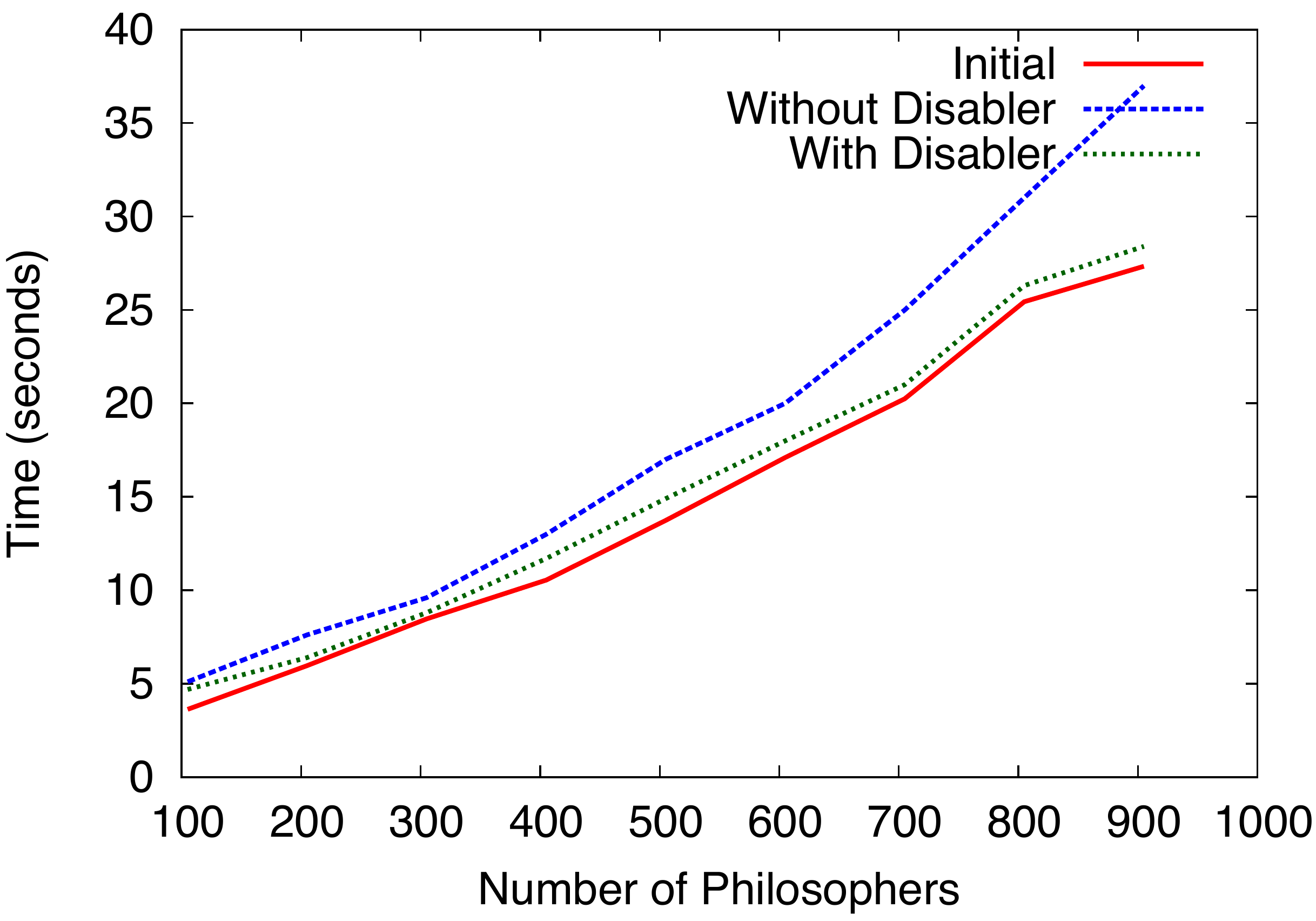}}
  {\includegraphics{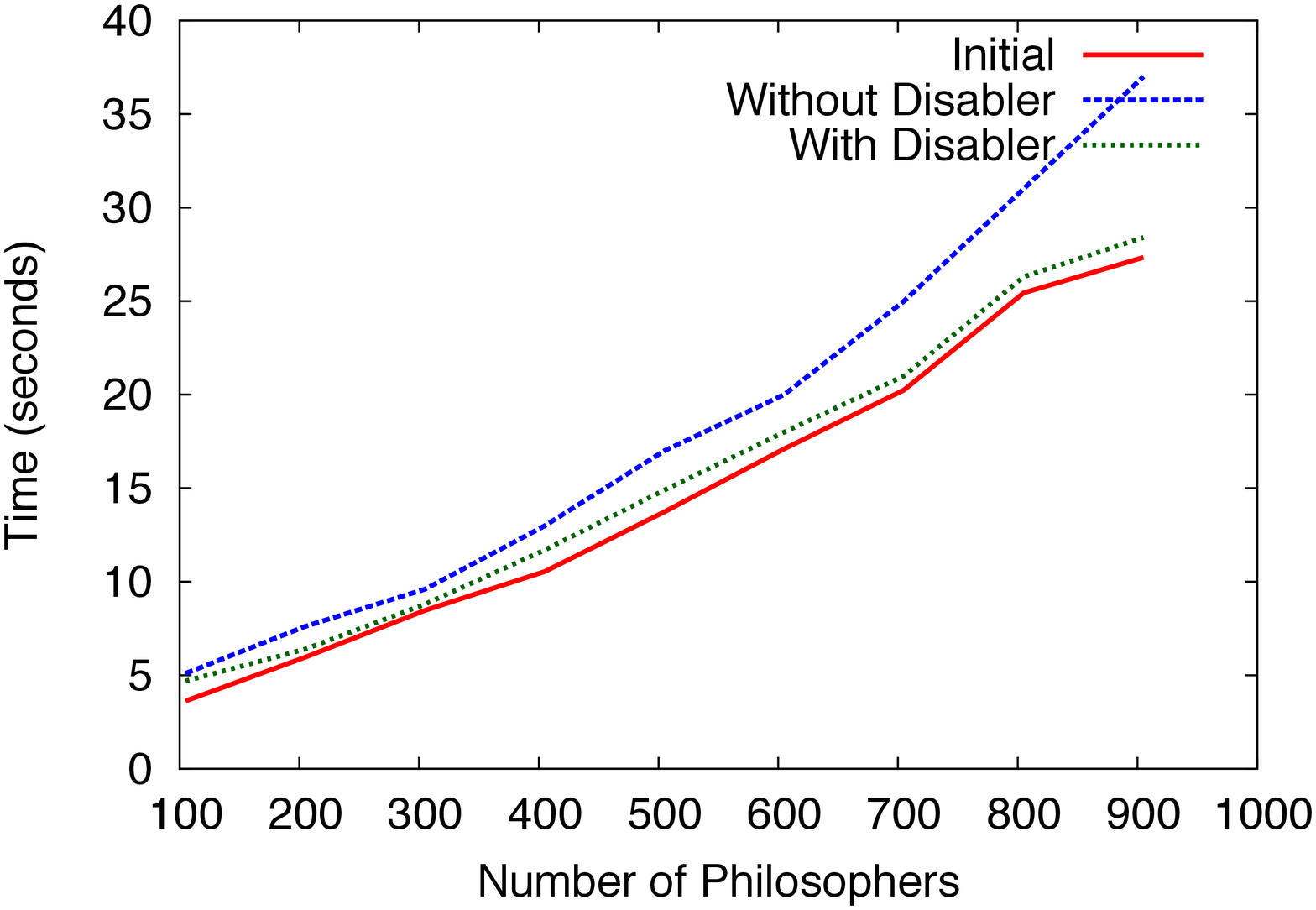}}
  }
  }
  \subfigure[Overhead of the supervised system]{
  \label{bench:overheadining}
  \scalebox{0.25}{
 \ifthenelse{\boolean{pdf}}
 {\includegraphics{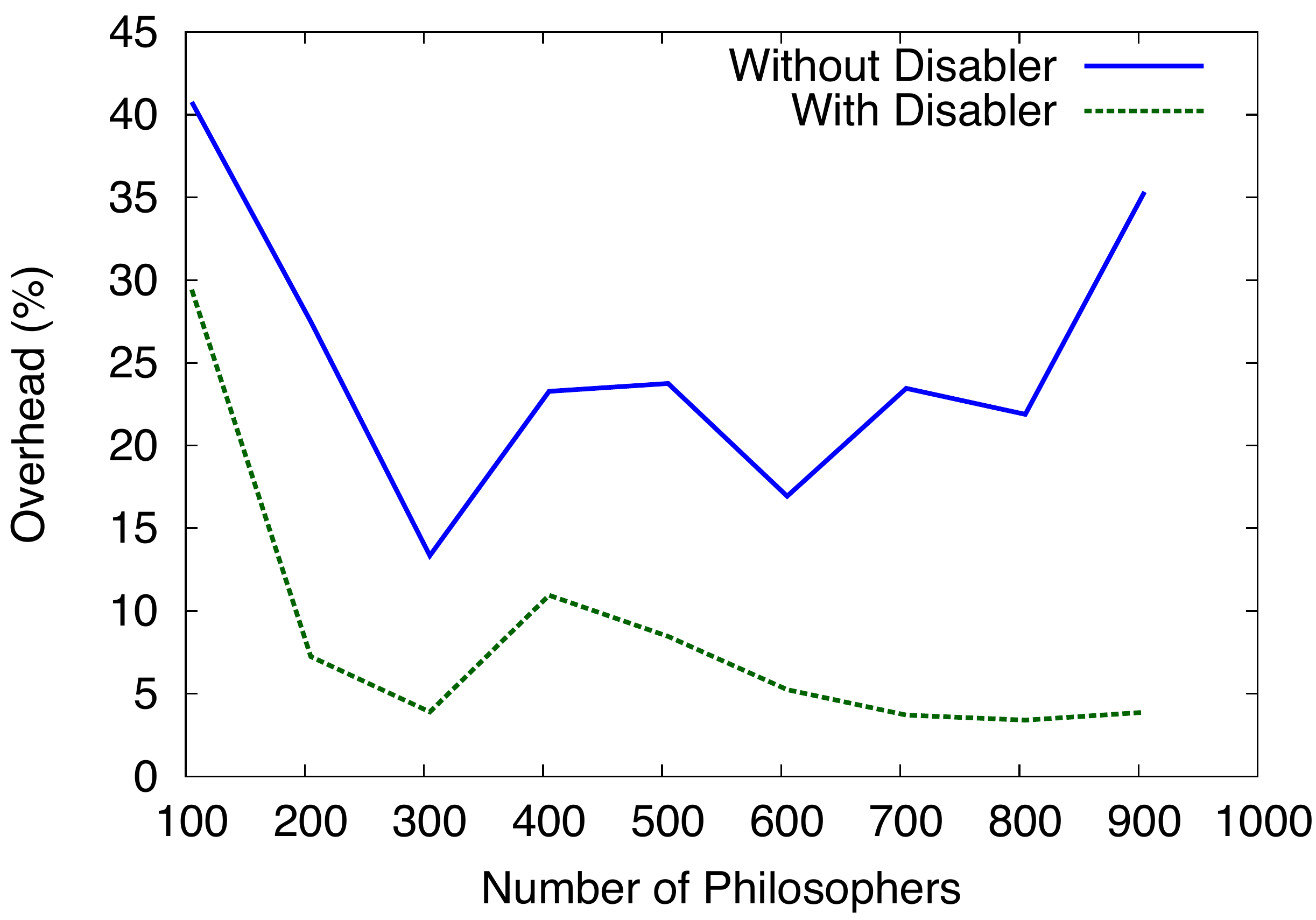}}
 {\includegraphics{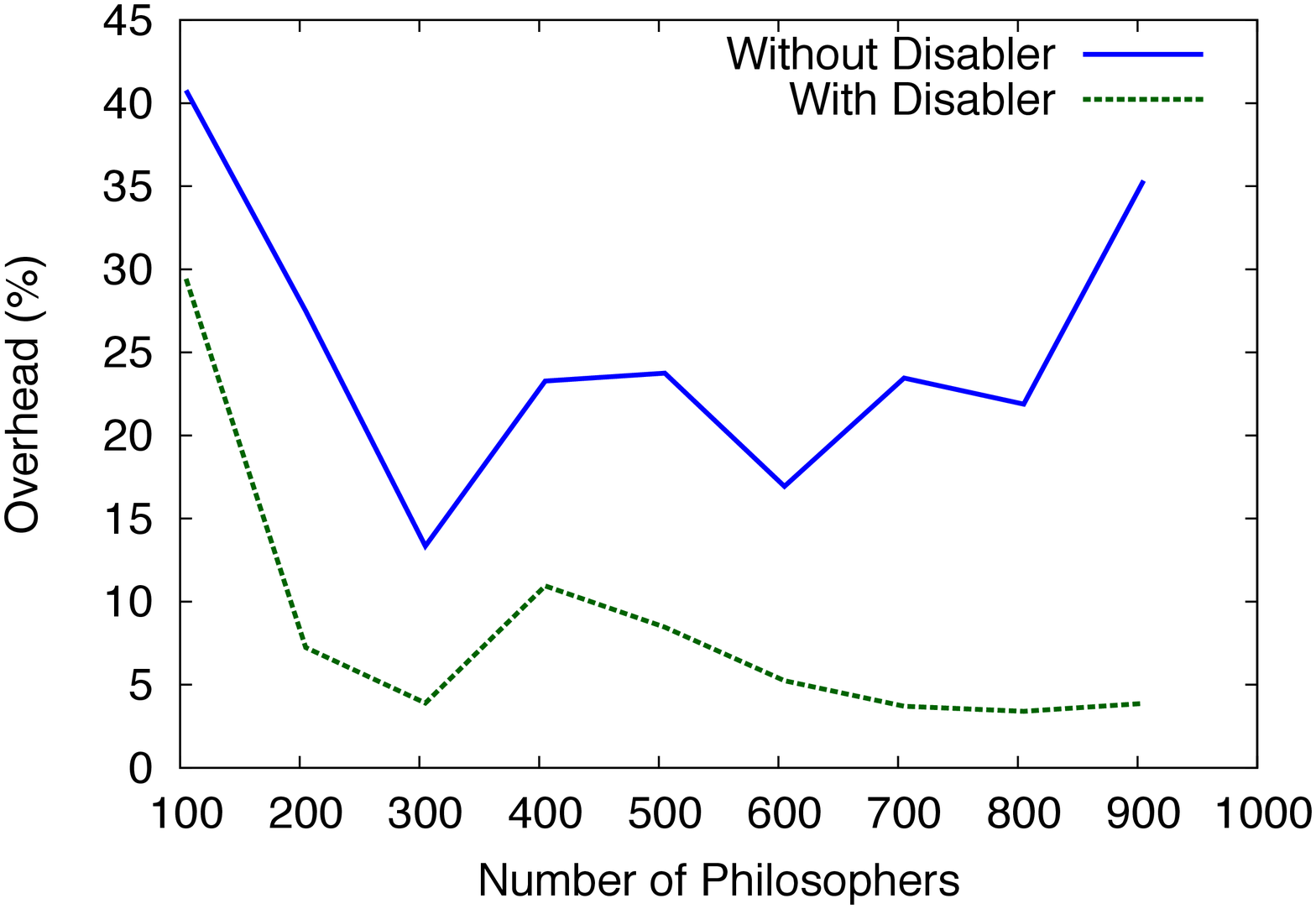}}
 }
 }
 \caption{Performance evaluation of dining philosophers}
 \label{fig:bench}
\end{figure}

We have modeled in BIP some dining philosophers that may deadlock.
We aim to enforce deadlock freedom at runtime.
Figure~\ref{fig:philo} (resp. \ref{fig:fork}) models the behavior of a philosopher (resp. fork) in BIP.
Figure~\ref{fig:dining} shows a composite system consisting of two philosophers and two forks. 
The system enters a deadlock state if all philosophers are in state $r$.
In that case, the system should recover. 
In Figures~\ref{bench:timedining}, \ref{bench:overheadining} we show some experimental results.
We increase the number of philosophers and compare the execution before and after the transformation (with and without disabler). 
The $x$-axis represents the number of philosophers (and also the number of forks). 
The $y$-axis represents the execution time. 
We ran the initial BIP system, which may deadlock, several times up to reach $10,000$ steps (i.e., $10,000$ releases of the fork).
We ran $10,000$ steps of the supervised BIP system (with and without the disabler). 
Our results show that the supervised system introduces a reasonable  overhead (e.g., $4\%$ in case of $900$ philosophers with disabler).
In this example, enabling the disabler, does not introduce deadlocks (there is always at least one good interaction after recovery, a philosopher with a fork on its right can take the fork on its left), and reduces significantly the overhead. 
%
\subsection{Using {\toolname} to Control Robots}
\label{sec:implem:robot}
%
\begin{figure}[t]
\centering
\scalebox{0.75}{
 \ifthenelse{\boolean{pdf}}
{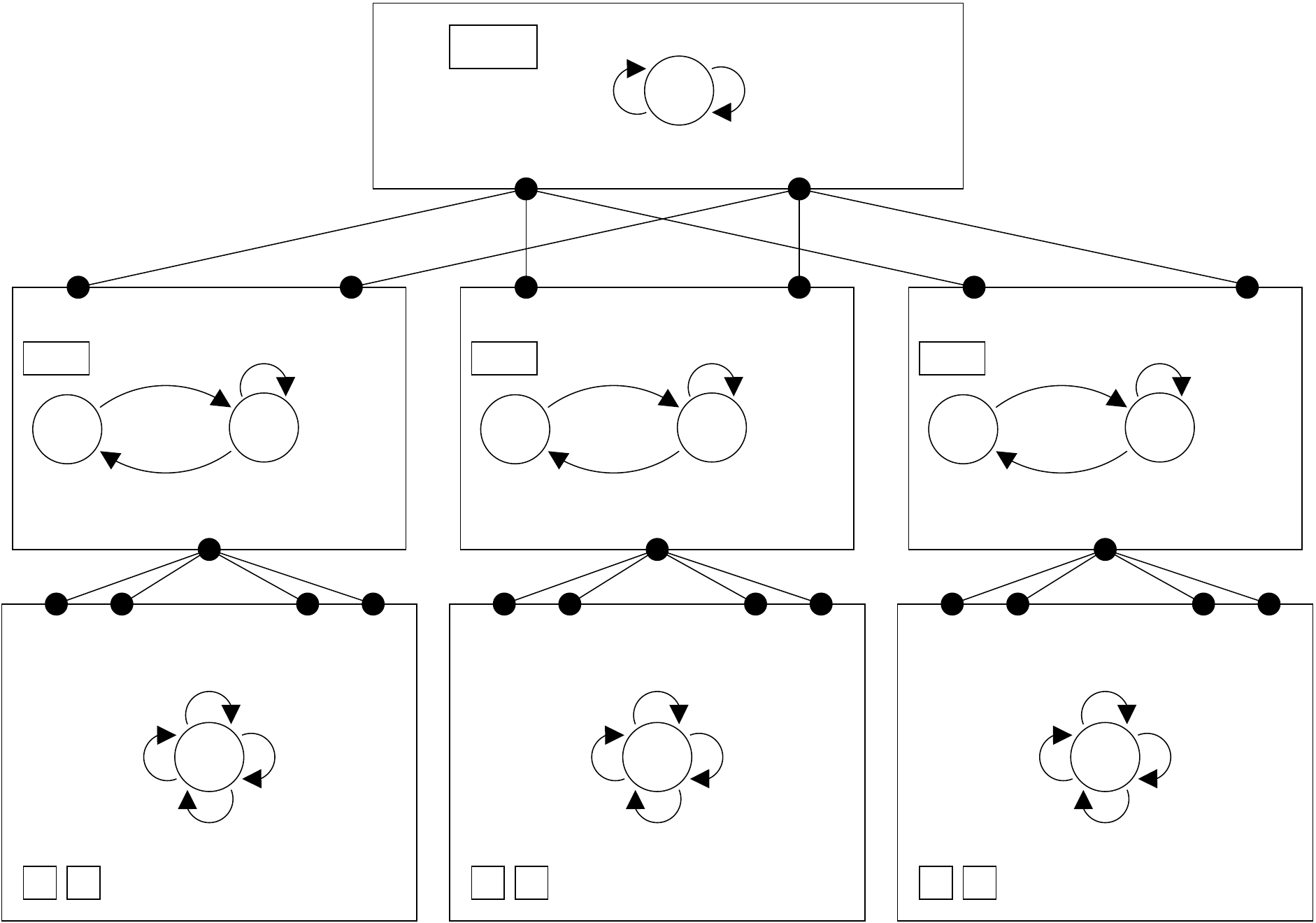}
{\input{fig/robot.pstex_t}}
}
\caption{Robotic application in BIP}
\label{fig:robots}
\end{figure}
Figure~\ref{fig:robots} shows a robotic system modeled in BIP.
We consider three robots (referred to as $R_i$, $i\in[1,3]$) placed on maps of size $n \times n$ with $n\in[2,5,100]$.
A robot can move up, down, left, and right. Each robot $R_i$ has a local controller $C_i$ that synchronizes with the robot to start and stop the robot.
When a robot starts, it randomly moves $1,000$ steps.
The system contains also a global controller $C$ that synchronizes with local controllers to count the number of active robots.
This model allows collisions between robots.
To avoid this,  the system must satisfy the following invariant\footnote{Invariants are stutter-invariant safety properties.} $\forall i, j \in [1,3]: R_i.x \neq R_j.x \vee R_i.y \neq R_j.y$.
Enforcing this invariant requires to manually modify the behaviors of robots as well as the architectures by adding new interactions.
This process is error-prone and the resuting system would be more complex and with reduced readability.

Using our method, we just create an oracle that emits verdict $\bot$ in case of collision of two robots (otherwise the verdict is $\top_c$), and the system is automatically instrumented to avoid collisions between robots.
This permits a separation of concerns between the main functionalities of the system and additional behaviors (e.g., avoiding collisions and ambush coordinates, limiting the number of active robots, etc.).  
Table~\ref{tab:bench:robots} shows the execution times (in seconds) to perform $2 \times 10^5$ correct (i.e., no collision) steps.
We generate four different configurations (Supervised, Supervised-d, Supervised-o, Supervised-o-d) of the supervised system.
We use -o to denote that the system is optimized, i.e., only the minimal set of transitions is instrumented.
We use -d to denote that a disabler has been integrated in the system.
For each configuration, we run the system on a map of different sizes ($n = 2, 5, 100$).
Obviously, the number of collisions decreases, and hence the number of rollbacks also decreases, with the size of the map.
For example, if we consider Supervised-o configuration and the map of size $n=2$ we obtain $400,280$ rollbacks and execution time (to perform $2 \times 10^5$ correct steps) $224$ seconds.
In this case, enabling the disabler (i.e., Supervised-o-d configuration) reduces the number of rollbacks and hence reduces the execution time ($177$ seconds).
Clearly, the optimized configurations outperform the non-optimized ones.
For maps of sizes $5$ and $100$ the disabler slightly reduces the number of collisions since the probability to take again the same step that has lead to a collision is very small.
Thus, in that case, enabling the disabler does not improve the execution time but adds a small overhead because of the interactions with the disabler.
\begin{table}[t]
\centering
\begin{tabular}{|c|c|c||c|c||c|c||c|c|}
\hline
\multirow{2}{*}{\backslashbox{Size ($n$)}{Configuration}} & \multicolumn{2}{|c||}{Supervised} & \multicolumn{2}{c||}{Supervised-d} & \multicolumn{2}{c||}{Supervised-o}  & \multicolumn{2}{c|}{Supervised-o-d} \\ \cline{2-9}
& \#Rollback & Time & \#Rollback & Time & \#Rollback & Time & \#Rollback & Time  \\ \hline \hline
2   & 399998 & 345  & 267001 &  282   & 400280  & 224 & 266549   & 177  \\ \hline
5   & 18039  & 129  & 16007  &  128   & 18022   & 82  & 15630    & 83 \\ \hline
100 & 68     & 122  & 53     &  120   & 35      & 76  & 50       & 78 \\ \hline
\end{tabular}
\caption{Execution times (seconds) and number of rollbacks of the BIP supervised robots}
\label{tab:bench:robots}
\end{table}
\section{Related Work}
\label{sec:rw}
%
\noindent
\textit{Model repair.}
Recent efforts (e.g., \cite{ChatzieleftheriouBSK12}) aim at adapting model-checking 
abstraction techniques to the model repair problem.
Our approach fundamentally differs from model repair  for several reasons.
First our approach operates at runtime: we do not statically modify the model of our system as our properties are expressive enough so that model-checking is undecidable or does not scale.
Moreover, our objective is to minimally alter the initial behavior of the system.
Correct executions in the initial system are preserved and yield observationaly equivalent executions in the supervised system.
\paragraph{Theories of fault-tolerance.}
The theory of fault-tolerance for CBSs was initiated by Arora and Kulkarni~\cite{AroraK98}.
Close to our approach is a framework for fault recovery in CBSs~\cite{BonakdarpourBG12}.
Bonakdarpour et al. assume a fault-model as input, i.e., a labelling of all transitions of the system as normal, faulty, and recovery, and then characterize the conditions for a system to converge to a normal behavior.
The authors target non-masking systems, i.e., systems i) where faults are recovered within a finite number of recovery actions, and ii) that always make progress.
Both the later and our approaches target BIP systems.
However, our approach has fundamental differences.
These approaches take as input fault-tolerant programs and assume fault-tolerance being clearly encoded inside the target program.
In~\cite{AroraK98}, the system is seen as a collection of guarded commands.
In~\cite{BonakdarpourBG12}, fault detection and recovery span across multiple components.
Both approaches fall short in meeting the modularity requirement of CBSs.
Indeed, programs in~\cite{AroraK98} do not have their own state-space.
The fault models considered in~\cite{BonakdarpourBG12} assumes fault detection and recovery to concern several components with inter-dependent interactions.
\paragraph{Supervisory approaches to fault-tolerance.}
Similar to our approach are techniques based on supervisory-control theory and controller synthesis \`a la Ramadge and Wonham~\cite{IntroDES}.
Similar objectives are to synthesize a mechanisms that is maximally permissive and ensures fault-tolerance by disabling the controllable transitions that would either make the system diverge from the expected behavior or prevent it from reaching the expected behavior.
In supervisory approaches the fault is due to a system action (cf.~\cite{WenKHL08}).
Faults are uncontrollable events and after their occurrence, the controller recovers the system within a finite number of steps. 
Moreover, the non-faulty part of the system needs to be available and distinguishable from the system.
Such approaches fall in the scope of our framework where monitors can enforce the non-occurrence of a particular action.
Moreover, as BIP systems usually contain data, guards and assignments, it is generally not possible to statically compute the faulty behavior in the system.
\paragraph{Runtime enforcement for monolithic systems.}
Several approaches were proposed for the runtime enforcement of (discrete-time) specifications over monolithic systems~(cf. \cite{enforceablesecpol,RuntimeNonSafety,FalconeMFR11}).
Several sets of enforceable properties were proposed with their associated enforcement monitors.
Restrictions to the set of enforceable specifications stem from the fact that the considered specifications are over infinite executions sequences.
As shown in~\cite{FalconeFM12}, when considering specifications over finite sequences, all properties become enforceable.
In this paper, we consider specifications over finite sequences but point out restrictions arising from the nature of the targeted systems.
It is also worth mentioning that the runtime enforcement paradigm proposed in this paper improves the previous ones.
Indeed, upon the detection of bad behaviors, previous enforcement paradigms proposed to ``accumulate events" in a memory (when dealing with progress properties) or halt the execution of the underlying system (when dealing with safety properties); with the hope that future events may help to satisfy the property again.
The enforcement paradigm proposed in this paper, studied now for safety properties but stated generally for any property, prevents and avoids the occurrence of faults by reverting the effect of events that lead to a deviation from the desired behaviors, leaving the system in a state just as before the fault occured.
\paragraph{Dynamic techniques for CBSs.}
Few dynamic approaches exist to improve the reliability of CBSs.
Dormoy et al. proposed FTPL, a customization of Linear Temporal Logic to specify the correctness of component reconfigurations in the Fractal framework~\cite{DormoyKL10}. Then, the authors proposed a runtime verification approach to the correctness of architectures~\cite{DormoyKL11}.
Independently, we proposed a runtime verification framework for BIP systems~\cite{FalconeJNBB13} that augments BIP systems with monitors for the conformance of the runtime behavior against linear-time properties.
All these approaches allowed only the \emph{detection} of errors and not their correction using recovery.
As the approach in~\cite{FalconeJNBB13} is only concerned with (the simpler problem) of runtime verification, it considers all properties as monitorable.
In this paper, we introduce a notion of enforceable properties specific to CBS and parametrized by a notion of number of tolerance steps.
While the purpose of the transformations in~\cite{FalconeJNBB13} is to introduce a monitor and transmit snapshots of the system to it, the transformations proposed in this paper additionally grant the monitor with primitives to backup the system state and control it.
As seen in \secref{sec:re-bip}, to preserve the consistency of the system in case of roll-back, not only the parts of the system involved with the property are instrumented but also the parts that are ``connected" to these.
\section{Conclusion and Future Work}
\label{sec:conclusion}
%
\paragraph{Conclusion.}
This paper introduces runtime enforcement for component-based systems described in the BIP framework.
Our approach considers an input system whose behavior may deviate from a desired specification.
We identify the set of stutter-invariant safety properties as enforceable on component-based systems.
Restrictions on the set of enforceable specifications come from i) the number of steps the system is allowed to deviate from the specification (before being corrected) and ii) the constraints imposed by instrumentation.
We propose a series of formal transformations of a (non-monitored) system to integrate an enforcement monitor, using the oracle of the specification as input.
Our validation approach is fully implemented in an available tool that has been used to enforce deadlock freedom on dining philosophers.
As a result, runtime enforcement provides an interesting complementary validation method as the validity of the specification is generally either undecidable or leads to an intractable state-explosion problem.
\paragraph{Some perspectives.}
In the future, we will consider more expressive properties (i.e., non-safety) such as $k$-step enforceable properties (with $k>1$) to allow transactional behavior.
It will entail to find an alternative instrumentation technique and avoid hard-coding the connections between the initial system and the monitor.
We will consider more dynamic connections between components using the (recent) dynamic version of BIP~\cite{BozgaJQS12}, combined with a memorization mechanism to store the state-history of components.

Moreover, we will work towards the decentralization of the enforcement monitor and the disabler to allow them to take decisions alone.
The expected benefit is to reduce communication in the system.
For this purpose, we shall inspire from~\cite{FalconeCF14} which considers the problem of decentralizing verification monitors in monolithic systems, and also from~\cite{BonakdarpourBJQS12} which distributes a centralized scheduler of components for a given distributed architecture.

Furthermore, we shall consider optimization techniques to further reduce the performance impact on the initial system.
For this purpose, we consider using static analysis on both the specification and the system to reduce the needed instrumentation.
%

%
\bibliographystyle{splncs}
\bibliography{biblio}

\begin{thebibliography}{10}

\bibitem{enforceablesecpol}
Schneider, F.B.:
\newblock Enforceable security policies.
\newblock ACM Trans. Inf. Syst. Secur. \textbf{3} (2000)  30--50

\bibitem{Falcone10}
Falcone, Y.:
\newblock You should better enforce than verify.
\newblock In Barringer, H., Falcone, Y., Finkbeiner, B., Havelund, K., Lee, I.,
  Pace, G.J., Rosu, G., Sokolsky, O., Tillmann, N., eds.: Proceedings of the
  1st International Conference on Runtime Verification (RV 10). Volume 6418 of
  Lecture Notes in Computer Science., Springer (2010)  89--105

\bibitem{FalconeMFR11}
Falcone, Y., Mounier, L., Fernandez, J.C., Richier, J.L.:
\newblock Runtime enforcement monitors: composition, synthesis, and enforcement
  abilities.
\newblock Formal Methods in System Design \textbf{38} (2011)  223--262

\bibitem{Bliudze-Sifakis-08a}
Bliudze, S., Sifakis, J.:
\newblock A notion of glue expressiveness for component-based systems.
\newblock In van Breugel, F., Chechik, M., eds.: Proceedings of the 19th
  International Conference on Concurrency Theory (CONCUR 2008). Volume 5201 of
  Lecture Notes in Computer Science., Springer (2008)  508--522

\bibitem{BliudzeS08}
Bliudze, S., Sifakis, J.:
\newblock The algebra of connectors---structuring interaction in {BIP}.
\newblock {IEEE} Transactions on Computers \textbf{57} (2008)  1315--1330

\bibitem{bip11}
Basu, A., Bensalem, S., Bozga, M., Combaz, J., Jaber, M., Nguyen, T.H.,
  Sifakis, J.:
\newblock {Rigorous Component-Based System Design Using the BIP Framework}.
\newblock IEEE Software \textbf{28} (2011)  41--48

\bibitem{LeuckerBS08JLC}
Bauer, A., Leucker, M., Schallhart, C.:
\newblock Comparing {LTL} semantics for runtime verification.
\newblock Journal of Logic and Computation \textbf{20} (2010)  651--674

\bibitem{DBLP:conf/rv/FalconeFM09}
Falcone, Y., Fernandez, J.C., Mounier, L.:
\newblock Runtime verification of safety-progress properties.
\newblock In Bensalem, S., Peled, D., eds.: Proceedings of the 9th
  International Workshop on Runtime Verification (RV 2009), Selected Papers.
  Volume 5779 of Lecture Notes in Computer Science., Springer (2009)  40--59

\bibitem{FalconeJNBB13}
Falcone, Y., Jaber, M., Nguyen, T.H., Bozga, M., Bensalem, S.:
\newblock Runtime verification of component-based systems in the {BIP}
  framework with formally proved sound and complete instrumentation.
\newblock SOftware and SYstem Modeling (2013) To appear. Pre-print available
  online.

\bibitem{RuntimeNonSafety}
Ligatti, J., Bauer, L., Walker, D.:
\newblock Run-time enforcement of nonsafety policies.
\newblock ACM Trans. Inf. Syst. Secur. \textbf{12} (2009)  19:1--19:41

\bibitem{FalconeFM12}
Falcone, Y., Fernandez, J.C., Mounier, L.:
\newblock What can you verify and enforce at runtime?
\newblock Software Tools for Technology Transfer \textbf{14} (2012)  349--382

\bibitem{Lamport83}
Lamport, L.:
\newblock What good is temporal logic?
\newblock In: IFIP Congress. (1983)  657--668

\bibitem{WilkeTemporal}
Wilke, T.:
\newblock Classifying discrete temporal properties.
\newblock In Meinel, C., Tison, S., eds.: Proceedings of the 16th Annual
  Symposium on Theoretical Aspects of Computer Science (STACS 99). Volume 1563
  of Lecture Notes in Computer Science., Springer (1999)  32--46

\bibitem{ChatzieleftheriouBSK12}
Chatzieleftheriou, G., Bonakdarpour, B., Smolka, S.A., Katsaros, P.:
\newblock Abstract model repair.
\newblock In Goodloe, A., Person, S., eds.: Proceedings of the 4th
  International Symposium on NASA Formal Methods (NFM 2012). Volume 7226 of
  Lecture Notes in Computer Science., Springer (2012)  341--355

\bibitem{AroraK98}
Arora, A., Kulkarni, S.S.:
\newblock Detectors and correctors: A theory of fault-tolerance components.
\newblock In: ICDCS. (1998)  436--443

\bibitem{BonakdarpourBG12}
Bonakdarpour, B., Bozga, M., G{\"o}{\ss}ler, G.:
\newblock A theory of fault recovery for component-based models.
\newblock In Richa, A.W., Scheideler, C., eds.: Proceedings of the 14th
  International Symposium on Stabilization, Safety, and Security of Distributed
  Systems (SSS 2012). Volume 7596 of Lecture Notes in Computer Science.,
  Springer (2012)  314--328

\bibitem{IntroDES}
Cassandras, C.G., Lafortune, S.:
\newblock Introduction to Discrete Event Systems.
\newblock Springer-Verlag, Secaucus, NJ, USA (2006)

\bibitem{WenKHL08}
Wen, Q., Kumar, R., Huang, J., Liu, H.:
\newblock A framework for fault-tolerant control of discrete event systems.
\newblock IEEE Trans. Automat. Contr. \textbf{53} (2008)  1839--1849

\bibitem{DormoyKL10}
Dormoy, J., Kouchnarenko, O., Lanoix, A.:
\newblock Using temporal logic for dynamic reconfigurations of components.
\newblock In Barbosa, L.S., Lumpe, M., eds.: Proceedings of the 7th
  International Workshop on Formal Aspects of Component Software (FACS 2010).
  Volume 6921 of Lecture Notes in Computer Science., Springer (2010)  200--217

\bibitem{DormoyKL11}
Dormoy, J., Kouchnarenko, O., Lanoix, A.:
\newblock Runtime verification of temporal patterns for dynamic
  reconfigurations of components.
\newblock In Arbab, F., {\"O}lveczky, P.C., eds.: Proceedings of the 8th
  International Symposium on Formal Aspects of Component Software, Revised
  Selected Papers, (FACS 2011). Volume 7253 of Lecture Notes in Computer
  Science., Springer (2011)  115--132

\bibitem{BozgaJQS12}
Bozga, M., Jaber, M., Maris, N., Sifakis., J.:
\newblock Modeling dynamic architectures using {Dy-BIP}.
\newblock In Gschwind, T., Paoli, F.D., Gruhn, V., Book, M., eds.: Proceedings
  of the 11th International Conference on Software Composition (SC 2012).
  Volume 7306 of Lecture Notes in Computer Science., Springer (2012)  1--16

\bibitem{FalconeCF14}
Falcone, Y., Cornebize, T., Fernandez, J.C.:
\newblock Efficient and generalized decentralized monitoring of regular
  languages.
\newblock In {\'A}brah{\'a}m, E., Palamidessi, C., eds.: FORTE 2014:
  Proceedings of the 34th IFIP WG 6.1 International Conference on Formal
  Techniques for Distributed Objects, Components, and Systems,. Volume 8461 of
  Lecture Notes in Computer Science., Springer (2014)  66--83

\bibitem{BonakdarpourBJQS12}
Bonakdarpour, B., Bozga, M., Jaber, M., Quilbeuf, J., Sifakis, J.:
\newblock A framework for automated distributed implementation of
  component-based models.
\newblock Distributed Computing \textbf{25} (2012)  383--409

\end{thebibliography}
\end{document}